\documentclass{article}
\usepackage{PRIMEarxiv}
\usepackage{amsmath}
\usepackage{amsfonts}
\usepackage{bm}
\usepackage[utf8]{inputenc} 
\usepackage[T1]{fontenc}    
\usepackage{hyperref}       
\usepackage{url}            
\usepackage{booktabs}       
\usepackage{amsfonts}       
\usepackage{nicefrac}       
\usepackage{microtype}      
\usepackage{lipsum}
\usepackage{fancyhdr}       
\usepackage{graphicx}       
\usepackage{multirow} 
\graphicspath{{media/}}     

\title{Learning Global and Local Features of Power Load Series Through Transformer and 2D-CNN: An image-based Multi-step Forecasting Approach Incorporating Phase Space Reconstruction
\thanks{This work was supported in part by the National Natural Science Foundation of China (51977082) and the Key-Area Research and Development Program of Guangdong Province (No. 2023B0909040003).} 
}

\author{
  Zihan Tang, Tianyao Ji, Wenhu Tang \\
  South China University of Technology \\
  \texttt{hayesscut@gmail.com, \{tyji, wenhutang\}@scut.edu.cn} \\
}

\begin{document}
\maketitle

\begin{abstract}
As modern power systems continue to evolve, accurate power load forecasting remains a critical
issue in energy management. The phase space reconstruction (PSR) method can effectively retain the inner chaotic property of power load from a system dynamics perspective and thus is a promising knowledge-based preprocessing method for power load forecasting. In order to fully utilize the capability of PSR method to model the non-stationary characteristics within power load, and to solve the problem of the difficulty in applying traditional PSR prediction methods to form a general multi-step forecasting scheme, this study proposes a novel multi-step forecasting approach by delicately integrating the PSR with neural networks to establish an end-to-end learning system. Firstly, the useful features in the phase trajectory obtained from the preprocessing of PSR are discussed in detail. Through mathematical derivation, the
equivalent characterization of the PSR and another time series preprocessing method, patch segmentation (PS), is demonstrated for the first time. Based on this prior knowledge, an image-based modeling
perspective with the global and local feature extraction strategy is introduced. Subsequently, a novel
deep learning model, namely PSR-GALIEN, is designed for end-to-end processing, in which the Transformer Encoder and 2D-convolutional neural networks (CNN) are employed for the extraction of the global and local patterns in the image, and a multi-layer perception (MLP) based predictor is used for the efficient correlation modeling. Then, extensive experiments are conducted on five real-world benchmark
datasets to verify the effectiveness as well as to have an insight into the detailed performance of the PSR-GALIEN. The results show that, compared with six state-of-the-art deep learning models, the
forecasting performance of PSR-GALIEN consistently surpasses these baselines, achieving superior
accuracy in both intra-day and day-ahead forecasting scenarios. At the same time, the attributions of its forecasting results can be explained through the visualization-based method, which significantly increases the interpretability.
\end{abstract}


\section{Introduction}\label{}

In modern power systems, the electricity consumption characteristics presented by end-users are constantly changing, as the increasing number of new loads and distributed renewable energy generation facilities has brought brand new challenges to the accurate power load forecasting, which is always the cornerstone of control and decision. The mathematical problem inherent in power load forecasting can be attributed to non-stationary time series forecasting \cite{alhussein2020hybrid}, and the field of time series forecasting is a vibrant research area which has grown considerably within the last few decades. In recent years, deep learning models has made great strides for their powerful nonlinear feature extraction capabilities, however, the one-fits-all approach is still far away from the existing techniques \cite{khodayar2020deep}. In order to weaken the negative effects of non-stationarity and thus enhance the adaptability and robustness of the predictor, feature engineering for preprocessing power load series is a crucial part through out the whole forecasting piplines, relevant methods of time series preprocessing can be divided into two mainstreams, namely, the decomposition methods and the embedding methods \cite{fulcher2018feature}.

Decomposition methods in signal processing are widely used for manually extracting different levels of components in the original time series, under the framework of decomposition-prediction-aggregation, there are numerous studies focusing on combining different decomposition algorithms with machine learning and deep learning models, investigating the predictive performance of these combinations \cite{yan2021frequency,bedi2020energy,hafeez2021novel,el2020ensemble,jnr2021hybrid,peng2021daily,huang2022multivariate}. Although such studies illustrate that these decomposition methods can effectively separate the different components which characterize the non-stationary characteristics of the power load series hence improving the predictability of the original series, and by mean of which good prediction accuracy can be achieved through the powerful learning-based predictors, they inevitably suffer from the following inherent problems, such as difficulties in implementation, high computational complexity, long inference time, heavy dependence on manual experience, etc. Moreover, these approaches fail to explore the source of the non-stationarity of the power load series.

Time-delay embedding, a classical time series analysis method, based on which the phase space reconstruction (PSR) theory can effectively uncover the inner nonlinear characteristics inherent in the time series by reconstructing them into the high-dimensional phase space to approximate its original dynamic structure \cite{kennel1992determining,ma2006selection,han2015wind}. For power load series, the unique advantage of phase space reconstruction is that it provides a way to explore the inner chaotic features from the perspective of nonlinear dynamics through using the observed sequences itself to reconstruct the inner structure, therefore reducing the heavy dependence on exogenous variables, which has several benefits in practical applications. Firstly, redundant computational overheads and additional prediction errors can be avoided. The exogenous features such as regional meteorological features are high-dimensional with redundancy, and usually there is a issue of time granularity mismatch, which result in heavy burden in data processing. What's worse, the quality of meteorological data is also dependent on meteorological forecasts, and inaccurate forecasts in correlation modelling may introduce additional errors in power load forecasting. Secondly, PSR provides a modeling perspective in phase space instead of in the time domain under the autoregressive scheme, which is more beneficial in the subsequent feature extraction, especially for the other scenarios where these data are not available for practical reasons, or the cost of obtaining them is unaffordable \cite{zjavka2016short,keyinformation2022,fan2018short}. 

Based on these advantages, PSR is widely used in the forecasting scenarios of wind speed, renewable energy generation, as well as power load consumption in power systems\cite{ji2022short,li2020midterm,wang2017multi,xu2023natural,sun2018short}, however, drawbacks still exist in the current research studies, an important fact, as argued in \cite{li2020midterm}, is that limited by the fundamental forecasting theory in phase space, most of the existing studies integrating prediction models with PSR have only investigated their single-step forecasting performance, as for the multi-step forecasting methods, the rolling forecast strategy is widely used, which generates cumulative errors. For example, literature \cite{fan2018short} adopts the conventional local forecasting method in the PSR theory to consider the spatial relationship between the neighboring phase points in the phase space, and proposes a single-step prediction model with a bi-square kernel regression approach for short-term power load forecasting. Literature \cite{shi2019phase} proposes an ultra-short-term forecasting model integrating the constrained Boltzmann machine predictor with PSR, namely PSR-DBN, and investigates its forecasting performance within the range of 12 steps ahead (1 hour) on a bus load dataset, literature \cite{huang2022short} proposes a solution by using the structure of long short-term memory (LSTM) network with fully-connected networks (FCN) for multi-step forecasting, however, the LSTM network still belongs to the single-step iterative processing scheme with slow inference speed and cumulative error, literature \cite{hou2022load} proposes a rolling data strategy to obtain a longer forecasting horizon, utilizing the ensemble learning with the LSTM networks for prediction, which is still the same idea. 

Overall, currently the vast majority of existing studies incorporating the PSR method fail to explore a general long-term multi-step forecasting methodology, in which an end-to-end correlation modeling methodology from the reconstructed phase trajectories to the time series to be predicted is believed to be a necessary part. This research gap hinders their performance in effectively addressing the scenarios where long range multi-step forecasts are needed, for example, in the real-world day-ahead load forecasting tasks, the forecasting horizon in advance is usually 48 steps or 96 steps, according to practical requirements for time granularity. This paper argues that there are mainly two reasons corresponding for this gap: (1) from the perspective of feature engineering, these studies fail to take full advantage of the data structure obtained from the preprocessing method of PSR and extract meaningful features which characterize certain properties of the phase trajectories and thus is beneficial to build correlation models from relative features to the power load series to be predicted.
(2) from the perspective of algorithm design, these studies fail to delicately integrate the PSR preprocessing method with the specific functional deep learning modules to form an efficient and powerful end-to-end pipeline including the preprocessing, feature extraction, and forecasting stages.

In order to address the above issues, this paper focuses on the following aspects, i.e., what useful features can be extracted from the phase trajectories, what kind of neural networks can be used to extract these features, and most importantly, how to design a subtle blend of the PSR preprocessing method and the deep-learning models to form such an unified scheme. In summary, The main contributions in this paper are:
\begin{enumerate}
\item A novel image-based modeling and forecasting framework is proposed to make full utilization of the PSR method in feature engineering, which adopts a global and local feature extraction strategy to simultaneously extract the corresponding patterns in the phase trajectory image, and their correlations to the future target series are established jointly. 

\item  Based on this framework, a novel deep-learning model, namely PSR-GALIEN, is proposed for efficient end-to-end processing, in which the Transformer Encoder and 2D-convolutional neural network (CNN) are implemented to extract the global and local features in this image respectively, and a multi-layer perception (MLP) based predictor is used for correlation modeling.

\item Extensive experiments are conducted to have a comprehensive insight to PSR-GALIEN’s performance, demonstrating the effectiveness of this image-based approach, as well as the superior forecasting accuracy of PSR-GALIEN in the real-world intra-day and day-ahead forecasting scenarios.

\item The feature interpretation method of using the regression activation map (RAM) and heatmap to explain the utilized features of PSR-GALIEN in the forecasting process is proposed, which is critical for understanding the attribution of the predictions results.
\end{enumerate}

The subsequent sections of this paper are organized as follows. In Section \ref{section 2}, the innovative image-based modeling idea as well as the related feature extraction method are introduced, and based on this framework, the design and implementation details of the proposed PSR-GALIEN model is described in section \ref{section 3}. In section \ref{section 4}, the experiments conducted in our research are introduced and the results are analysed. Lastly, in section \ref{section 5} the conclusion and future works are summarized.
\section{Proposed image-based methodology}
\label{section 2}
\subsection{Preliminaries in Phase space reconstruction}
\subsubsection{Principle}
In the classical theory of PSR, the perspective of system dynamics is used to analyze the nonlinear properties of chaos in time series, which considers the chaotic time series as the projection of the attractors in high-dimensional space into one-dimensional space \cite{gao2009complex}. By means of time-delay embedding, the original time series can be reconstructed into phase trajectories in high-dimensional space to approximate its dominant attractor. The theory proposed by Takens \cite{noakes1991takens}, Whitney \cite{whitney1936differentiable}, et al., mathematically guarantees that this embedding is a diffeomorphic representation of the original nonlinear system with proper reconstruction parameters setup, i.e, the delay time $\tau$, and embedding dimension $m$. Without loss of generality, given a set of observed time series $[{x_1},{x_2},\ldots,{x_t}]$, the embedding dimension $m$ and the time delay $\tau$, the embedding phase trajectories can be obtained from the trajectory matrix, as described in Eq. \ref{PSR embedding equa}. 

\begin{equation}
\begin{array}{l}
{\bf{T}} = \left[ {\begin{array}{*{20}{c}}
{{{\bf{x}}_1}}& \ldots &{{{\bf{x}}_i}}& \ldots &{{{\bf{x}}_N}}
\end{array}} \right] = \left[ {\begin{array}{*{20}{c}}
{{x_1}}& \ldots &{{x_i}}& \ldots &{{x_N}}\\
 \vdots &{}& \vdots &{}& \vdots \\
{{x_{1 + j\tau }}}& \ldots &{{x_{i + j\tau }}}& \ldots &{{x_{N + j\tau }}}\\
 \vdots &{}& \vdots &{}& \vdots \\
{{x_{1 + (m - 1)\tau }}}& \ldots &{{x_{i + (m - 1)\tau }}}& \ldots &{{x_{N + (m - 1)\tau }}}
\end{array}} \right]
\end{array}
\label{PSR embedding equa}
\end{equation}
where $\mathbf{x}_i \in \mathbb{R}^{m \times 1}$ is the $i$-th phase point in the trajectory $[\mathbf{x}_1,\mathbf{x}_2,\ldots,\mathbf{x}_{N}]$.


Each point on the phase trajectory can be viewed as a state in a certain nonlinear system, the time-delay embedding technique greatly enriches the semantic information of original time series which can be considered as the observation of the system, for this reason, capturing the dynamic features among these phase points in the whole trajectory allows for a better way to analyse the nonlinear roots inherent in the orginal time series. 

Like wind speed series, photovoltaic (PV) series, and other time series that are strongly coupled to the earth-atmosphere system, power load series are considered to have the chaotic properties to some extend, which is the source of their nonlinear characteristics \cite{xiong2021blended,chen2022online,wang2020improved}. The largest Lyapunov exponent (LLE) is a widely acknowledged indicator for examining and quantifying the chaotic characteristics of time series, if the LLE is positive, then the intrinsic dynamical system that dominates the time series has chaotic properties. In this paper, the Wolf algorithm \cite{WOLF1985285} is used to calculate LLE of each dataset, which evaluates the degree of chaos of the sequence by calculating the separation speed of as the two neighboring orbits located in the phase space.

\subsubsection{Parameter selection}
The proper selection of reconstruction parameters is crucial to uncover and approximate the nonlinear characteristics of the intrinsic system in time series. Parameter selection can be performed through multiple methods, which are categorized into two types: the individual estimation methods and the joint estimation methods. In our study, as two classical analytical methods for chaotic time series, the mutual information (MI) method \cite{PMID:9896728} and the false nearest neighbor (FNN) \cite{PhysRevA.45.3403} method are implemented to estimate the delay time and the embedding dimension respectively.

The mutual information method uses the mutual information function (MIF) between the original sequence $[{x_1},{x_2},...,{x_t}]$ and its delayed sequence $[{x_{1 + \tau }},{x_{2 + \tau }},...,{x_{t + \tau }}]$ for delay time estimation, and its optimal delay time is the global minimal of the mutual information function, at which the original sequence has the lowest degree of nonlinear correlation with all of its own delayed sequences. Without loss of generality, given two random variables $X$ and $Y$, their mutual information is:

\begin{equation}
    I(X,Y) = \sum\limits_{x,y} {p(x,y)\log \frac{{p(x,y)}}{{p(x)p(y)}}}
\end{equation}
The optimal estimation of the delay time is:
\begin{equation}
{\tau ^*} = \arg \mathop {\min }\limits_\tau  \sum\limits_{{x_t},{x_{t + \tau }}}^{} {p({x_t},{x_{t + \tau }})} \log \frac{{p({x_t},{x_{t + \tau }})}}{{p({x_t})p({x_{t + \tau }})}}
\label{calculation tau}
\end{equation}

For the determination of the embedding dimension, Takens et al. \cite{noakes1991takens} mathematically gives the theoretical basis that the embedding dimension should satisfy $m$>$2d+1$,where $d$ is the fractal dimension of the attractor. The idea of estimating the embedding dimension by applying the false neighbor point method is that when the phase trajectories are not sufficiently expanded, the points that are originally far apart in the high dimensional space may become the neighbors of each other when they are projected into the low dimensional space. As the embedding dimension keeps increasing, the gradual unfolding of the collapsed phase trajectory will cause the percentage of these false neighboring points to keep decreasing, and then, in this case, the original attractor structure can be fully characterized and described. Considering the reconstructed phase trajectories in the d-dimensional phase space, for arbitary phase point $\mathbf{x}_i(d)$, and its the geometrically nearest neighbor in Euclidean space is $\mathbf{x}_i^n(d)$, given the threshold $\varepsilon$, the false nearest neighboring relationship is satisfied when:
\begin{equation}
\frac{{{{\left\| {{\mathbf{x}_i}(d + 1) - \mathbf{x}_i^n(d + 1)} \right\|}^2}}}{{{{\left\| {{\mathbf{x}_i}(d) - \mathbf{x}_i^n(d)} \right\|}^2}}} \ge \varepsilon ,i = 1,2,3,...
\label{calculation m}
\end{equation}
By examining the mapping between the number of these points and the embedding dimension, an appropriate estimation can be made, for example, one feasible way of estimating the optimal embedding dimension is the case when the proportion of all the phase points possessing such a false nearest neighboring relationship is less than 5\%.

\subsection{Image-based modeling and forecasting framework}
\begin{figure*}[h!]
    \centering
\includegraphics[width=\textwidth]{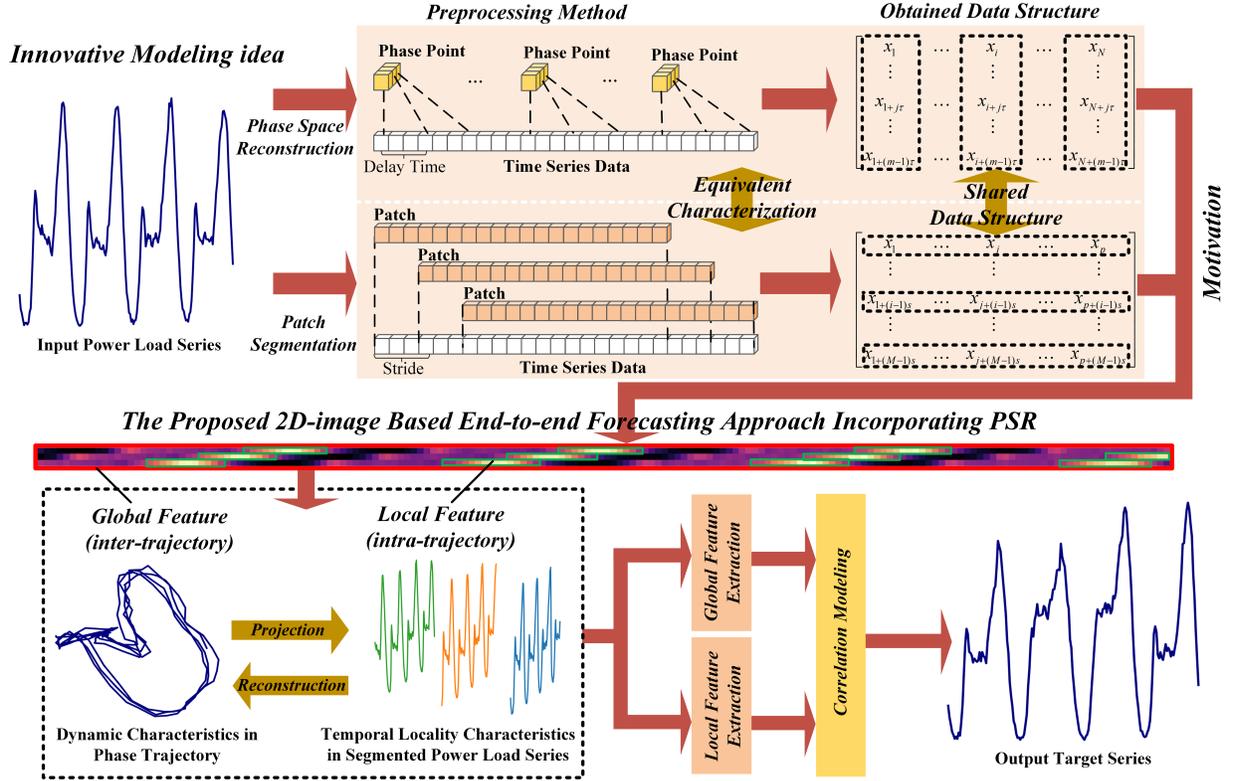}
    \caption{Image-based feature extraction and forecasting framework incorporating the PSR preprocessing method}
    \label{data structure}
\end{figure*}
\subsubsection{General Learning-based forecasting paradigm of PSR}
Traditional forecasting theory based on PSR mainly includes two mainstream methods: the global prediction methods and the local prediction methods, where global prediction methods use the evolutionary trajectories of all the phase points in the phase space to fit the nonlinear dynamical relations, and local prediction methods are concerned with the use of similar phase points in the historical evolutionary trajectories to guide for the future. Despite being supported by mathematical basis, these two methods can not effectively address the scenario of multi-step forecasting.

Deep learning models are capable of extracting the evolutionary laws in nonlinear dynamical systems automatically, which makes it possible to integrate them with these conventional prediction methods and thus to extend their predictive performance through establishing the direct nonlinear mapping relationships. Without loss of generality, under the autoregressive forecasting scheme, the general prediction model based on this modeling approach can be expressed by the following equation:

\begin{equation}
{\bf{y}}_{t:t + \omega }^T = f([{{\bf{x}}_1},{{\bf{x}}_2},{{\bf{x}}_3},...,{{\bf{x}}_N}])
\label{forecasting psr}
\end{equation}
where ${\bf{y}}_{t:t + \omega }^T = [{x_{t + 1}},{x_{t + 2}},...,{x_{t + \omega }}]$, $\omega$ is the forecasting horizon. $[{{\bf{x}}_1},{{\bf{x}}_2},{{\bf{x}}_3},...,{{\bf{x}}_N}]$ represents the phase trajectory, and $f( \cdot )$ represents its end-to-end nonlinear correlation with the future time series.

\subsubsection{Data structure and  feature analysis}
In the learning-based forecasting model, having a clear perception of the features among the data structure of input is essential for the subsequent network design. From the trajectory matrix $\mathbf{T}$, it is clear that the column of the matrix represent a phase point in the trajectory, while the row of the matrix represent the projection sequences in a certain dimension. It is demonstrated in this study that the process of obtaining these projection sequences can be regarded as another important method in time series feature engineering: the patch segmentation (PS) preprocessing method, which means that after being processed by both methods, the same data structure can be obtained, and more importantly, this also implies that a methodology can be adopted to fully exploit the corresponding features in this shared data structure under the two different perspective for the downstream correlation modelling, with a view to improving the final forecasting accuracy.

The feature engineering method of PS has been widely used in the tasks of image processing \cite{dosovitskiy2020image}, text processing \cite{DBLP:journals/corr/abs-1810-04805}, speech signal processing \cite{lea2017temporal}, etc. In the time series forecasting tasks, the PS method brings an approach for patch-level correlation modeling through extracting the local patterns in the segmented subseries, and compared to the most widely used point-level correlation modeling approach, this modelling approach is more advantageous because patches contain richer semantic information from which a short-term local evolutionary pattern of a time series can be extracted and utilized.

Consider the process of PS in terms of data structures, the same with PSR, two parameters are required for reshaping the original time series into the specified form , i.e., the length of the segmented patch sub-series $p$, and the length of the stride between each patches $s$. Once the segmentation parameters are determined, the data structure of the original time series $[{x_1},{x_2},...,{x_t}]$ will be reorganized as described in the following equation:

\begin{equation}
\begin{array}{l}
{\bf{P}} = \left[ {\begin{array}{*{20}{c}}
{{\bf{p}}_1^T}\\
 \vdots \\
{{\bf{p}}_i^T}\\
 \vdots \\
{{\bf{p}}_M^T}
\end{array}} \right] =
\left[ {\begin{array}{*{20}{c}}
{{x_1}}&{...}&{{x_j}}&{...}&{{x_p}}\\
 \vdots &{}& \vdots &{}& \vdots \\
{{x_{1 + (i - 1)s}}}&{...}&{{x_{j + (i - 1)s}}}&{...}&{{x_{p + (i - 1)s}}}\\
 \vdots &{}& \vdots &{}& \vdots \\
{{x_{1 + (M - 1)s}}}&{...}&{{x_{j + (M - 1)s}}}&{...}&{{x_{p + (M - 1)s}}}
\end{array}} \right]
\end{array}
\label{patch segentation equa}
\end{equation}
where ${{\bf{p}}_i^T}\in \mathbb{R}{^{{1} \times p}}$ is the $i$-th segmented subseries from the time series. From the two matrices in Eq. \ref{PSR embedding equa} and Eq. \ref{patch segentation equa}, the mathematical connection between the two data preprocessing methods can be found:
\begin{equation}
{\bf{P}} = {\bf{T}},{\rm{ when }}\left\{ {\begin{array}{*{20}{c}}
{\tau  = s}\\
{t - (m - 1)\tau  = p}
\end{array}} \right.
\label{equivalence proof}
\end{equation}

Without loss of generality, considering the non-uniform time-delay embedding based phase space reconstruction method used in \cite{du2024novel,han2018nonuniform}. i.e., the delay times are written in a more universal form:

\begin{equation}
\begin{array}{l}
{\bf{T}} = [{{\bf{x}}_1}, \ldots ,{{\bf{x}}_i}, \ldots ,{{\bf{x}}_N}]= \left[ {\begin{array}{*{20}{c}}
{{x_1}}& \ldots &{{x_i}}& \ldots &{{x_N}}\\
 \vdots &{}& \vdots &{}& \vdots \\
{{x_{1 + \sum\limits_{j - 1} \tau  }}}& \ldots &{{x_{i + \sum\limits_{j - 1} \tau  }}}& \ldots &{{x_{N + \sum\limits_{j - 1} \tau  }}}\\
 \vdots &{}& \vdots &{}& \vdots \\
{{x_{1 + \sum\limits_{m - 1} \tau  }}}& \ldots &{{x_{i + \sum\limits_{m - 1} \tau  }}}& \ldots &{{x_{N + \sum\limits_{m - 1} \tau  }}}
\end{array}} \right]
\end{array}
\end{equation}
where $\sum\limits_{j - 1} \tau   = \sum\limits_{i = 1}^{j - 1} {{\tau _i}}$, at this point, Eq. \ref{equivalence proof} is rewritten as:


\begin{equation}
{\bf{P}} = {\bf{T}},{\rm{when}}\left\{ {\begin{array}{*{20}{c}}
{{\bm{\tau }} = {\bf{s}}}\\
{t - \sum\limits_{j = 1}^{m - 1} {{\tau _j} = p} }
\end{array}} \right.
\label{constrain}
\end{equation}
where ${\bm{\tau }} = {[{\tau _1}, {\tau _2},{\tau _3},...,{\tau _{M-1}}]^T}$, and the corresponding stride vector is ${\bf{s}} = {[{s_1},{s_2},...,{s_{M - 1}}]^T}$.

From the above derivation, we can conclude that when the condition of Eq. \ref{constrain} is satisfied, the matrices obtained by the two preprocessing methods are equivalent. However, despite sharing the same data structure, the meaning of the features inherent in it is quite different under the modeling perspectives of the two different approaches, as illustrated in Fig. \ref{data structure}. 

It should be noted that the trade-offs need to be made to consider the way in which this shared data structure is obtained, as the parameters used in the process have practical significance, and decision needs to be given to the prominent characteristics of the time series. For power loads, we mainly consider how to use the method of PSR to extract the chaotic features in it, and at the same time, we can use the PS method to extract the temporal locality features from the project sequences in order to bring additional information gain for the forecasting, which we believe is a knowledge gap in the previous studies.

\subsubsection{Feature extraction and Correlation modeling strategy}
In the previous subsections, the data structure and inner features of the phase trajectory is discussed in detail. Thus, considering the coupling and spatial relations of the two different features under the shared matrix data structure, in order to be more flexible in extracting these features from the phase trajectory, and drawing on the feature engineering approaches from the vision Transformer (ViT) \cite{dosovitskiy2020image} and Conformer \cite{peng2021conformer}, this paper discards the conventional sequential modelling method for feature extraction and treats it as a two-dimensional greyscale image. At the same time, the global and local feature extraction strategy is adopted for the extracting the corresponding features analysed in the previous subsections, which are defined as:
\begin{itemize}
    \item \textbf{Global Feature}: The overall evolution characteristics of phase trajectory in m-dimensional phase space, which can be characterized by the global pattern of the image, as shown in the red box in Fig. \ref{data structure}.
    \item \textbf{Local Feature}: The temporal locality characteristics of the projection sequences of the phase trajectory in each dimension (equivalently, the segmented patch subseries), which can be characterized by the local patterns of the image, as shown in the green boxes in Fig. \ref{data structure}.
\end{itemize}

Taking the above analyses together, it is now clear what kind of features can be extracted from the phase trajectories, intuitively, both the two features can be extracted and utilized to establish a correlation model, but the connection between them is unrealised. Thus, based on this prior knowledge which is demonstrated in the subsections, we consider to form a jointly correlation model to make the best of this data structure.

In the next section, the way to build a deep learning architecture to conduct such feature extraction as well as feature correlation strategy to form an end-to-end learning system will be introduced.

\section{Proposed model architecture}
\label{section 3}
\begin{figure*}[h!]
    \centering
    \includegraphics[width=\textwidth]{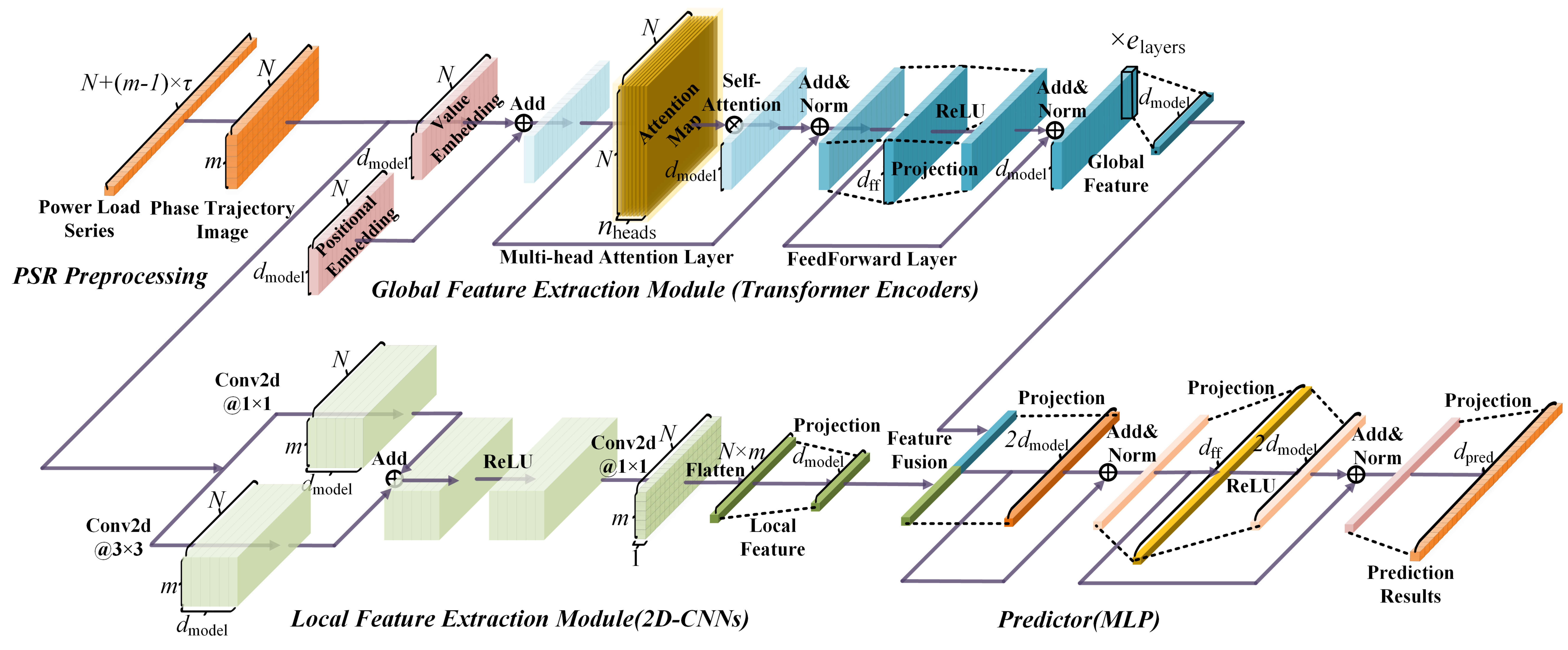}
    \caption{Architecture of the proposed PSR-GALIEN model, where $n_{\rm{heads}}$,$e_{\rm{layers}}$, $d_{\rm{model}}$, and $d_{\rm{ff}}$ are hyperparameters representing the number of heads in the multi-head attention layer of Transformer Encoder,the number of implemented Encoder layers, the number of hidden neurons in the feature extraction network and in the MLP predictor, respectively.}
    \label{model-picture}
\end{figure*}

In order to realize the above modeling methodology and make full utilization of deep learning models, the proposed architecture in this paper consists of a two-branch feature extraction network that uses the Transformer Encoder and 2D-CNNs to capture the global and local features respectively, the global and local features extracted from the two modules are combined and then feed into the MLP predictor, and the final prediction results are obtained. As the priori knowledge, the process of PSR is included in the preprocessing stage, which is well integrated with the latter networks to form the learning system. The proposed deep learning architecture is named as \textbf{P}hase \textbf{S}pace \textbf{R}econstruction based \textbf{G}lobal \textbf{A}nd \textbf{L}ocal \textbf{I}nformation \textbf{E}nhanced neural \textbf{N}etwork (\textbf{PSR-GALIEN}), as shown in Fig. \ref{model-picture}. The design and implementation details of each component of the network are introduced in the following subsections.

\subsection{Transformer-based global feature extraction module}
The Transformer model, which is centered on the self-attention mechanism, can efficiently establish the long-range dependencies among the input sequence, under the proposed image-based modeling perspective, considering each phase point in a phase trajectory as a image patch, the self-attention mechanism in Transformer Encoder can be used to establish a spatial correlation model between all the patches, and thus through which a global feature representing the overall evolution characteristics of the whole trajectory is obtained. 

In the global feature extraction module of the PSR-GALIEN model, the image data obtained from the PSR first needs the embedding process including the value embedding (VE) and the positional embedding (PE) before entering into the Transformer Encoder. VE is implemented by using a fully-connected layer, which aims to learn the useful representation of the image patches, and PE is designed to bring additional position-informed feature after the VE, in order to solve the problem of position-independent nature of the self-attention mechanism. After the processing of the embedding layer, the patch sequence with the length of $N$ and the embedding dimension of ${d_{{\rm{model}}}}$ enters into the Transformer Encoder layer for the further processing, where the multi-head attention layer and the feed-forward layer in it are at the core of nonlinear correlation modeling of these image patches. Given input sequence ${\bf{X}} \in \mathbb{R}{^{{d_{{\rm{model}}}} \times N}}$, the calculation process of the self-attention mechanism can be expressed by the following equations:
\begin{equation}
\begin{array}{l}
{{\bf{Q}}_i} = {\bf{W}}_q^i{\bf{X}}, 
 {{\bf{K}}_i} = {\bf{W}}_k^i{\bf{X}}, 
 {{\bf{V}}_i} = {\bf{W}}_v^i{\bf{X}}
\end{array}
\label{QKV}
\end{equation}

\begin{equation}
{{\bf{A}}_i} = {\rm{Softmax}}\left( {\frac{{{\bf{K}}_i^T{{\bf{Q}}_i}}}{{\sqrt {{d_{{\rm{model}}}}} }}} \right)
\end{equation}

\begin{equation}
\begin{array}{cc}
     &  \\
     & 

{\bf{X}}_i^{{\rm{output}}} = {\rm{SelfAtten}}\left( {{{\bf{X}}_i}} \right) = {\bf{X}}{{\bf{A}}_i}
\end{array}
\end{equation}
where ${\bf{Q}}_{i}\in \mathbb{R}{^{{d_{{\rm{model}}}} \times N}}$, ${\bf{K}}_{i}\in \mathbb{R}{^{{d_{{\rm{model}}}} \times N}}$, ${\bf{V}}_{i}\in \mathbb{R}{^{{d_{{\rm{model}}}} \times N}}$ are the query vectors, key vectors and value vectors respectively. ${\bf{A}}_{i} \in \mathbb{R}{^{{N} \times N}}$ is the attention matrix, which can be visualized see how much attention the model pays to each location within the sequence, and ${\bf{X}}_i^{{\rm{output}}} \in \mathbb{R}{^{{d_{{\rm{model}}}} \times N}}$ is the output sequence, $i$ denotes the $i$-th self-attention head. 

For the multi-head attention mechanism, it use multiple heads to obtain different representations of the patterns in order to improve the modeling capability, which can be expressed as:
\begin{equation}
{\rm{MultiAtten}}\left( {\bf{X}} \right) = {\bf{W}} \cdot {\rm{Concatenate}}({\bf{X}}_1^{{\rm{output}}}, \ldots ,{\bf{X}}_{{n_{{\rm{heads}}}}}^{{\rm{output}}})
\end{equation}
where ${\rm{Concatenate}}( \cdot )$ represent the processing of matrix concatenation by column, and ${\bf{W}}$ is the projection matrix for feature fusion.

It should be noted that the only parameters that can be learned in the Multi-head attention layer are the matrices used for linear transformations as demonstrated in Eq. \ref{QKV}, and the correlation modeling relies on the parameter-free inner product operation, thus a fully-connected layer with nonlinear activation functions is needed for modeling the nonlinear relations within the sequence, and the subsequent process is:

\begin{equation}
\begin{array}{l}
{\bf{Y}} = {\rm{LN}}\left( \begin{array}{l}
{\rm{LN}}\left( {{\bf{X}} + {\rm{MultiAtten}}\left( {\bf{X}} \right)} \right) + \\
{\rm{FFN}}\left( {{\rm{LN}}\left( {{\bf{X}} + {\rm{MultiAtten}}\left( {\bf{X}} \right)} \right)} \right)
\end{array} \right)\\
{\rm{ = [}}{{\bf{y}}_1}{\rm{,}}{{\bf{y}}_2}{\rm{,}}...{\rm{,}}{{\bf{y}}_N}{\rm{]}}
\end{array}
\end{equation}
where ${{\bf{Y}}} \in \mathbb{R}{^{{d_{{\rm{model}}}} \times N}}$ represent the output sequence of the Transformer Encoder. ${\rm{MultiAtten}}(\cdot)$, ${\rm{LN}}\left( {\cdot} \right)$, ${\rm{FFN}}\left( {\cdot} \right)$ denote the process performed by the multi-head attention layer, the layer normalization module and the fully-connected layer respectively.

The global feature extraction module in PSR-GALIEN employs $e_{\rm{layers}}$ Enocoder layers in series for feature extraction, and the last image patch (the most recent phase point connected to the time series to be predicted) in the output sequence possesses all the information of the previous patches, thus it is considered as the global representation of the whole trajectory. The global feature ${{\bf{y}}_{\rm{g}}}\in \mathbb{R}{^{{d_{{\rm{model}}}} \times N}}$ is finally obtained through a linear transform, as expressed in:
\begin{equation}
{{\bf{y}}_{\rm{g}}} = {{\bf{W}}_{\rm{g}}}{{\bf{y}}_N}
\end{equation}
where ${{{\bf{W}}_l}}\in \mathbb{R}{^{{{d_{{\rm{model}}}}} \times {{d_{{\rm{model}}}}}}}$ is the learnable matrix for obtaining the global feature.
\subsection{CNN-based local feature extraction module}
For the design of the feature extraction network for local patterns of the phase trajectory image, considering the huge gap between the length and the width of this image (determined by Eq. \ref{PSR embedding equa}, \ref{calculation tau} and \ref{calculation m}, usually $N \gg m$.), the conventional modeling method of using CNNs with deep structure for a large receptive field (RF), such as ResNet-50, VGG-16, may not be so beneficial in this case for the following reasons:
\begin{enumerate}
\item As illustrated in the previous subsections, the Transformer Encoders are already used to capture the global features in the image, and there is no need to use another deep CNN to account for the long-range correlation. In fact, the local feature extraction branch based on CNN is implemented to extract the detailed local patterns of the image (e.g., the local temporal patterns in each projection sequences of phase trajectory), the two networks are designed to undertake different feature extraction tasks as complementary. 

\item As the network continues to deepen, the unbalanced aspect ratio in the image makes it more difficult for the deeper layers to obtain effective semantic features, because most of the contents in their RF are the values obtained from padding, and these parts do not have any meaningful information, but significantly increase the computational overhead instead, and more importantly, a CNN with too complicated deep structure may hinder the efficient training of it.
\end{enumerate}
Overall, instead of using deep structures, a wide structure with multi-scale perception CNNs will make more sense in this case, like the the classical Inception network proposed in \cite{szegedy2015going}. 

The local feature extraction network in the PSR-GALIEN model uses the modeling methodologies of a series of multi-scale perception CNNs for feature extraction. Firstly, two CNNs with the kernel size of 1*1 and 3*3 are implemented for the extraction of local patterns in the input image ${\bf{\Pi }}\in {\mathbb{R}^{1 \times m \times N}}$, and the feature maps with both ${d_{{\rm{model}}}}$ channels obtained from them are added up for feature fusion. Then, the integrated feature map is fed into a nonlinear layer based on the ReLU function to enhance the nonlinear representation capability of the previous CNN layer, and after that the channel compression is performed by a CNN with the kernel size of 1*1 to obtain the local feature map with 1 channel. Finally, the local feature ${{\bf{y}}_{\rm{l}}}\in \mathbb{R}{^{{d_{{\rm{model}}}} \times 1}}$ is obtained through the matrix flattening and linear projection. The whole processing of the local feature extraction module can be described by the following equations:

\begin{equation}
\begin{array}{l}
{\bf{\Pi }^{'}} = {\rm{Conv2d}}\left( {{\bf{\Pi }},{\rm{kernel\_size}} = 1,{\rm{padding}} = 0,{\rm{stride}} = 1} \right)\\
 + {\rm{Conv2d}}\left( {{\bf{\Pi }},{\rm{kernel\_size}} = 3,{\rm{padding}} = 1,{\rm{stride}} = 1} \right)
\end{array}
\end{equation}

\begin{equation}
\begin{array}{l}
{{\bf{\Pi }}^{''}} = {\rm{ReLU}}({{\bf{\Pi }}^{'}})
\end{array}
\end{equation}
\begin{equation}
{{\bf{\Pi }}^{'''}} = {\rm{Conv2d(}}{{\bf{\Pi }}^{''}},{\rm{kernel\_size}} = 1,{\rm{padding}} = 0,{\rm{stride}} = 1) 
\end{equation}

\begin{equation}
{{\bf{y}}_{\rm{l}}} = {{\bf{W}}_{\rm{l}}}{\rm{Flatten}}{({{\bf{\Pi }}^{'''}})^T}
\end{equation}
where ${\bf{\Pi }^{'}}\in {\mathbb{R}^{{d_{{\rm{model}}}} \times m \times N}}$, ${{\bf{\Pi }}^{''}}\in {\mathbb{R}^{{d_{{\rm{model}}}} \times m \times N}}$, ${{\bf{\Pi }}^{'''}}\in {\mathbb{R}^{{d_{{\rm{model}}}} \times m \times N}}$ stand for the feature map after feature fusion, the feature map after nonlinear activation and the obatained local feature map, respectively. ${\rm{Conv2d}}\left(  \cdot  \right)$, ${\rm{Flatten}}\left(  \cdot  \right)$ represent the operation of a 2D-CNN layer and matrix flattening, and ${{{\bf{W}}_l}}\in \mathbb{R}{^{{{d_{{\rm{model}}}}} \times {(m\times N)}}}$ is the learnable matrix for obtaining the local feature.

\subsection{MLP-based prediction module}
Taking the MLP network as the predictor can efficiently establish the nonlinear correlation between the extracted features and the target sequences, which retains all the information in the forecasting stage without any loss. Firstly, the global and local features in the phase trajectory image extracted by the two feature extraction modules are combined through vector concatenation, and then a three-layer feed-forward neural network was used to perform the feature transformation, in which both layers are structured with the residual connection to ensure that both linear and nonlinear representation can be learned in that process. Finally, the enhanced feature vector are output as the vector representing the time-series to be predicted through a linear transform. The prediction process above is expressed as:
\begin{equation}
y_{{\rm{proj}}}^1 = {\rm{LN}}\left( {{\bf{y}} + {{\bf{W}}_1}{\bf{y}}} \right)
\end{equation}
\begin{equation}
y_{{\rm{proj}}}^2 = {\rm{LN}}\left( {y_{{\rm{proj}}}^1 + {{\bf{W}}_3}{\rm{RELU}}\left( {{{\bf{W}}_2}y_{{\rm{proj}}}^1} \right)} \right)
\end{equation}

\begin{equation}
{y_{{\rm{out}}}} = {{\bf{W}}_4}y_{{\rm{proj}}}^2 + {\bf{b}}
\end{equation}
where ${{\bf{y}}_{{\rm{out}}}}\in \mathbb{R}{^{{d_{{\rm{pred}}}} \times 1}}$ is the vector of prediction results, ${{{\bf{W}}_1}}\in \mathbb{R}{^{{2d_{{\rm{model}}}} \times {2d_{{\rm{model}}}}}}$, ${{{\bf{W}}_2}}\in \mathbb{R}{^{{d_{{\rm{ff}}}} \times {2d_{{\rm{model}}}}}}$, ${{{\bf{W}}_3}}\in \mathbb{R}{^{{2d_{{\rm{model}}}} \times {d_{{\rm{ff}}}}}}$, ${{{\bf{W}}_4}}\in \mathbb{R}{^{{d_{{\rm{pred}}}} \times {2d_{{\rm{model}}}}}}$ represent the linear transformation matrices in each layer of the MLP predictor, and ${{\bf{b}}}\in \mathbb{R}{^{{d_{{\rm{pred}}}} \times {1}}}$ is the bias in the last layer.

\section{Experiments and analysis}
\label{section 4}
\subsection{Datasets}
In order to verify the effectiveness of the proposed PSR-GALIEN model in practical power load forecasting scenarios, five real-world power load datasets with varying characteristics are selected as the benchmark datasets for the experiments, which are described as follows:
\begin{itemize}
\item \textbf{Elia-2022}: the power load data from 2022 for the interconnected power grid of Elia, the Belgian power system operator.
\item \textbf{City-1}: the power load data from 2012 to 2014 for a distribution network in a city in southern China.
\item \textbf{Area-1 \& Area-2}: the power load data from 2009 to 2015 for two regions in southern China.
\item \textbf{Australia}: the residential power load data from 2013 for 300 customers aggregated in a community located in New South Wales, Australia.
\end{itemize}

\begin{figure*}[h!]
    \centering
\includegraphics[width=\textwidth]{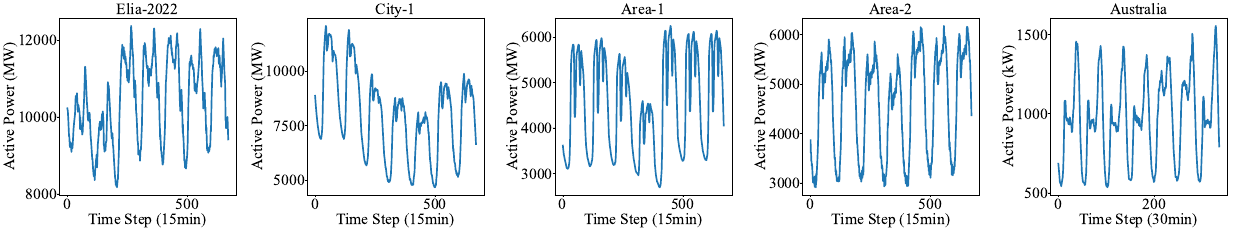}
    \caption{Visualization of the five datasets}
    \label{datasets}
\end{figure*}

These datasets cover scenarios with different spatial granularities ranging from aggregated residential users to large interconnected systems grids, which are well capable of verifying the robustness of the predictors among different scenarios. The detailed information of the datasets are shown in the Tab. \ref{dataset tab} and the temporal characteristics of each dataset are visualized in Fig. \ref{datasets}. As shown in the table, the LLE of each dataset is greater than 0, implying that the chaotic feature is a universal inner property among power loads despite of their varying temporal characteristics, and the estimated optimal reconstruction parameter of each dataset are presented.


\begin{table*}[h!]
  \centering
  \caption{The basic information of the 5 datasets in the experiments}
  \resizebox{\linewidth}{!}{
    \begin{tabular}{ccccccc}
    \toprule
    \rm{Dataset} & \rm{Time Granularity} & \rm{Training}\&\rm{Validation Set} & \rm{Test Set} & \rm{Total Set} &\rm{PSR Parameters} &\rm{LLE}\\
    \midrule
    \rm{Elia-2022} & \rm{15min} & \rm{28032 (292 days)} & \rm{7008 (73 days)} & \rm{35040 (365 days)} &  $\tau  = 40, m = 5$ & \rm{0.0175} 
    \\
    \rm{City-1} & \rm{15min} & \rm{52378 (546 days)} & \rm{13094 (136 days)} & \rm{65472 (682 days)} & $\tau  = 27, m = 4$ & \rm{0.0136} 
    \\
    \rm{Area-1} & \rm{15min} & \rm{169036 (1761 days)} & \rm{42260 (440 days)} & \rm{211296 (2201 days)} & $\tau  = 30, m = 4$ & \rm{0.0389}
    \\
    \rm{Area-2} & \rm{15min} & \rm{169036 (1761 days)} & \rm{42260 (440 days)} & \rm{211296 (2201 days)} & $\tau  = 28, m = 4$ & \rm{0.0213}
    \\
    \rm{Australia} & \rm{30min} & \rm{14016 (292 days)} & \rm{3504 (73 days)} & \rm{17520 (365 days)} & $\tau  = 8, m = 4$ & 
    \rm{0.0152}
    \\
    \bottomrule
    \end{tabular}%
    }
  \label{dataset tab}%
\end{table*}%
\subsection{Experimental setup}
\subsubsection{Baseline setup}
Deep learning models have made great progress in the field of time series forecasting in recent years, to keep up with the the latest trends as well as look back on the past milestones, six state-of-the-art and representative deep learning predictors are chosen as the baselines in the study, including Time Series Transformer (TST) \cite{vaswani2017attention}, Informer \cite{zhou2021informer}, PatchTST \cite{nie2022time} in Transformer family, DLinear \cite{zeng2023transformers}, Koopa \cite{liu2024koopa} in MLP family, and TimesNet \cite{wu2022timesnet} in CNN family. Among them, the TST model establishes a solid bridge between natural language processing and time series forecasting tasks, the Informer model extends its capability to perform long sequence forecasting, DLinear breaks the claim for the first time that the Transformer-structured models are more advantageous, as for the Koopa model, it employs the koopman operator to model the non-stationary properties in the time series from the perspective of nonlinear dynamics, which is relevant to the modeling idea of PSR. PatchTST and TimesNet are the two patch-based state-of-the-art models emerging in recent years, which are most relavent to our PSR-GALIEN. Therefore, these representative models were chosen as for comparison in this study, meanwhile, the performance of these state-of-the-arts on power load forecasting tasks is tested.

\subsubsection{Hyperparameter setup}

When conducting experiments, the six baselines are compared with the default settings presented in their original studies, as for the hyperparameters which are shared in the same structure, such as $d_{\rm{model}}$, $d_{\rm{ff}}$, $e_{\rm{layers}}$, $n_{\rm{heads}}$ in the Transformer backbone, are kept same, which are: $d_{\rm{model}}=512$, $d_{\rm{ff}}=4d_{\rm{model}}$, $e_{\rm{layers}}=2$, $n_{\rm{heads}}=8$ for Transformer, Informer, PatchTST and the PSR-GALIEN. As for the hyperparameters in the training stage, the setups are shown in Tab. \ref{hyper-table}.

\begin{table}[h!]
  \centering
  \caption{Hyperparameter setup in the training stage}
  \resizebox{0.4\linewidth}{!}{
    \begin{tabular}{cc}
    \toprule
    \rm{Hyperparameters} & \rm{Default configuration} \\
    \midrule
    \rm{Batch size} & \rm{32} \\
    \rm{Training epochs} & \rm{10} \\
    \rm{Learning rate} & \rm{0.0001} \\
    \rm{Look-back length} & \rm{96, 192, 336, 672} \\
    \rm{Forecasting length} & \rm{12 (6), 24 (12), 48 (24), 96 (48)} \\
    \rm{Random seeds} & \rm{2020, 2021, 2022, 2023, 2024} \\
    \rm{Loss function} & \rm{MSE} \\
    \rm{Optimizer} & \rm{Adam} \\
    \rm{Early stop counting} & \rm{3} \\
    \bottomrule
    \end{tabular}%
    }
  \label{hyper-table}%
\end{table}%

\subsubsection{Evaluation metrics}
In this paper, two main metrics in the time series forecasting task, the mean absolute error (MAE) and the mean absolute percentage error (MAPE)  are adopted for evaluation, which are fomulated by:
\begin{equation}
{\rm{MAE = }}\frac{1}{m}\sum\limits_{i = 1}^m {\left| {{y_i} - y_i^{{\rm{pred}}}} \right|}
\end{equation}

\begin{equation}
{\rm{MAPE = }}\frac{1}{m}\sum\limits_{i = 1}^m {\frac{{\left| {{y_i} - y_i^{{\rm{pred}}}} \right|}}{{{y_i}}}}  \times 100\% 
\end{equation}
where ${{y_i}}$, ${y_i^{{\rm{pred}}}}$ are the $i$-th value of the groundtruth and the prediction results respectively, $m$ is the prediction length, lower values of MAE and MAPE indicate higher prediction accuracy.

\subsubsection{Hardware and Software setup}
All the subsequent experiments in this study are carried out on a single machine with the hardware configuration of a NVIDIA RTX4090 GPU (24GB), a AMD EPYC 9754 vCPU (128-Core) and a RAM(24GB). All programs are implemented in the PyTorch 2.1.0-Python 3.10 (ubuntu22.04)-CUDA 12.1 environment.

\subsection{Experiments and analysis}
The experiments conducted in this paper consist of a total of four parts, firstly, the prediction performance of each model is compared on five datasets under the conditions of different forecast lengths. Secondly, the effects of varying input lengths on the prediction performance of each model are further explored on three long datasets, and then an ablation experiment is conducted to analyze the enhancement effects of the local feature extraction module in the PSR-GALIEN model. Finally, the impact of the various hyperparameters in PSR-GALIEN are discussed in detail.

\subsubsection{Comparison experiments with varying output lengths}
\begin{table*}[t]
\centering
\caption{Experimental results for different prediction lengths (P) on five datasets under fixed input length L = 192, with the best \textbf{bolded} and the second best \underline{underlined}.}
\resizebox{\linewidth}{!}{
\Large
\begin{tabular}{c|c|cc|cc|cc|cc|cc|cc|cc}
\toprule
\multicolumn{2}{c|}{\multirow{2}{*}{\rm{Model}}}
&\multicolumn{2}{c|}{\rm{PSR-GALIEN}}
&\multicolumn{2}{c|}{\rm{PatchTST} \cite{nie2022time}}
&\multicolumn{2}{c|}{\rm{Koopa} \cite{liu2024koopa}}
&\multicolumn{2}{c|}{\rm{DLinear} \cite{zeng2023transformers}}
&\multicolumn{2}{c|}{\rm{TimesNet} \cite{wu2022timesnet}}
&\multicolumn{2}{c|}{\rm{Informer} \cite{zhou2021informer}}
&\multicolumn{2}{c}{\rm{Transformer} \cite{vaswani2017attention}}

\\

\multicolumn{2}{c|}{} & \multicolumn{2}{c|}{\rm{Ours}} & \multicolumn{2}{c|}{\rm{2023}} & \multicolumn{2}{c|}{\rm{2024}} & \multicolumn{2}{c|}{\rm{2023}} & \multicolumn{2}{c|}{\rm{2023}} & \multicolumn{2}{c|}{\rm{2021}}& \multicolumn{2}{c}{\rm{2018}}\\
\hline
\multicolumn{2}{c|}{\rm{Metrics}} & \rm{MAE} & \rm{MAPE} &\rm{MAE} & \rm{MAPE} & \rm{MAE} & \rm{MAPE} & \rm{MAE} & \rm{MAPE} & \rm{MAE} & \rm{MAPE} & \rm{MAE} & \rm{MAPE} & \rm{MAE} & \rm{MAPE}\\ 
\hline

\multirow{4}{*}{\rotatebox{90}{\rm{Elia-2022}}} & \rm{12} 
&\pmb{\rm{148.146}} &\pmb{\rm{0.0164}} &\underline{\rm{168.872}} &\underline{\rm{0.0186}} 
&\rm{232.954} &\rm{0.0253} &\rm{198.107} &\rm{0.0218} 
&\rm{300.897} &\rm{0.0328} &\rm{213.906} &\rm{0.0237} 
&\rm{183.432} &\rm{0.0201} 

 \\

\multirow{4}{*}{} & \rm{24} 
&\pmb{\rm{201.027}} &\pmb{\rm{0.0222}} &\underline{\rm{210.578}} &\underline{\rm{0.0234}} 
&\rm{241.119} &\rm{0.0265} &\rm{255.828} &\rm{0.0282} 
&\rm{345.918} &\rm{0.0375} &\rm{302.971} &\rm{0.0327} 
&\rm{256.640} &\rm{0.0281}

\\

\multirow{4}{*}{} & \rm{48}
&\pmb{\rm{259.438}} &\pmb{\rm{0.0285}} &\underline{\rm{267.353}} &\underline{\rm{0.0305}} 
&\rm{322.626} &\rm{0.0352} &\rm{324.442} &\rm{0.0357} 
&\rm{388.725} &\rm{0.0422} &\rm{375.116} &\rm{0.0406} 
&\rm{370.911} &\rm{0.0401}

\\

\multirow{4}{*}{} & \rm{96}
&\pmb{\rm{325.566}} &\pmb{\rm{0.0357}} &\underline{\rm{350.906}} &\underline{\rm{0.0387}} 
&\rm{397.991} &\rm{0.0433} &\rm{399.379} &\rm{0.0438} 
&\rm{443.665} &\rm{0.0479} &\rm{427.972} &\rm{0.0466} 
&\rm{416.789} &\rm{0.0459} 

\\
\hline

\multirow{4}{*}{\rotatebox{90}{\rm{City-1}}} & \rm{12}
&\pmb{\rm{104.267}} &\pmb{\rm{0.0146}} &\underline{\rm{144.416}} &\underline{\rm{0.0204}} 
&\rm{343.308} &\rm{0.0473} &\rm{159.395} &\rm{0.0220} 
&\rm{256.480} &\rm{0.0363} &\rm{177.477} &\rm{0.0258} 
&\rm{176.485} &\rm{0.0271} 

\\

\multirow{4}{*}{} & \rm{24}
&\pmb{\rm{136.989}} &\pmb{\rm{0.0188}} &\underline{\rm{178.718}} &\underline{\rm{0.0253}} 
&\rm{192.794} &\rm{0.0266} &\rm{206.058} &\rm{0.0285} 
&\rm{287.225} &\rm{0.0405} &\rm{240.345} &\rm{0.0345} 
&\rm{243.637} &\rm{0.0387} 
\\

\multirow{4}{*}{} & \rm{48} 
&\pmb{\rm{184.181}} &\pmb{\rm{0.0250}} &\underline{\rm{207.207}} &\underline{\rm{0.0282}} 
&\rm{257.460} &\rm{0.0349} &\rm{271.851} &\rm{0.0371} 
&\rm{345.247} &\rm{0.0485} &\rm{276.334} &\rm{0.0393} 
&\rm{318.431} &\rm{0.0497} 
 
\\

\multirow{4}{*}{} & \rm{96} 
&\pmb{\rm{229.314}} &\pmb{\rm{0.0309}}&\underline{\rm{259.560}} &\underline{\rm{0.0352}} 
&\rm{335.406} &\rm{0.0452} &\rm{336.695} &\rm{0.0462} 
&\rm{365.023} &\rm{0.0502} &\rm{340.111} &\rm{0.0487} 
&\rm{396.053} &\rm{0.0594} 

\\
\hline

\multirow{4}{*}{\rotatebox{90}{\rm{Area-1}}} & \rm{12}
&\pmb{\rm{59.701}} &\pmb{\rm{0.0086}} &\underline{\rm{97.864}} &\underline{\rm{0.0144}} 
&\rm{296.494} &\rm{0.0444} &\rm{163.159} &\rm{0.0242} 
&\rm{162.604} &\rm{0.0243} &\rm{134.639} &\rm{0.0191} 
&\rm{129.798} &\rm{0.0199} 

\\

\multirow{4}{*}{} & \rm{24} 
&\pmb{\rm{90.116}} &\pmb{\rm{0.0131}} &\underline{\rm{121.122}} &\underline{\rm{0.0179}} 
&\rm{181.112} &\rm{0.0263} &\rm{227.659} &\rm{0.0338} 
&\rm{198.604} &\rm{0.0298} &\rm{162.417} &\rm{0.0234} 
&\rm{164.403} &\rm{0.0260}

\\

\multirow{4}{*}{} & \rm{48} 
&\pmb{\rm{131.221}} &\pmb{\rm{0.0191}} &\underline{\rm{160.687}} &\underline{\rm{0.0237}} 
&\rm{211.226} &\rm{0.0313} &\rm{325.451} &\rm{0.0481} 
&\rm{252.738} &\rm{0.0382} &\rm{208.980} &\rm{0.0303} 
&\rm{213.115} &\rm{0.0331}

\\

\multirow{4}{*}{} & \rm{96} 
&\pmb{\rm{194.516}} &\pmb{\rm{0.0289}} &\underline{\rm{221.934}} &\underline{\rm{0.0334}} 
&\rm{303.440} &\rm{0.0458} &\rm{452.536} &\rm{0.0678} 
&\rm{317.123} &\rm{0.0491} &\rm{298.012} &\rm{0.0452} 
&\rm{301.976} &\rm{0.0469} 

\\
\hline

\multirow{4}{*}{\rotatebox{90}{\rm{Area-2}}} & \rm{12}
&\pmb{\rm{87.591}} &\pmb{\rm{0.0117}} &\underline{\rm{112.559}} &\underline{\rm{0.0155}} 
&\rm{236.010} &\rm{0.0312} &\rm{142.064} &\rm{0.0190} 
&\rm{161.773} &\rm{0.0217} &\rm{151.333} &\rm{0.0194} 
&\rm{127.899} &\rm{0.0167} 

\\

\multirow{4}{*}{} & \rm{24}
&\pmb{\rm{117.406}} &\pmb{\rm{0.0153}} &\underline{\rm{139.294}} &\underline{\rm{0.0185}} 
&\rm{176.564} &\rm{0.0232} &\rm{193.373} &\rm{0.0254} 
&\rm{198.427} &\rm{0.0265} &\rm{201.354} &\rm{0.0255} 
&\rm{177.896} &\rm{0.0239} 

\\

\multirow{4}{*}{} & \rm{48}
&\pmb{\rm{159.145}} &\pmb{\rm{0.0203}} &\underline{\rm{179.204}} &\underline{\rm{0.0231}} 
&\rm{235.555} &\rm{0.0303} &\rm{265.646} &\rm{0.0342} 
&\rm{241.169} &\rm{0.0318} &\rm{266.833} &\rm{0.0335} 
&\rm{242.157} &\rm{0.0325} 

\\

\multirow{4}{*}{} & \rm{96} 
&\pmb{\rm{223.444}} &\pmb{\rm{0.0283}} &\underline{\rm{237.260}} &\underline{\rm{0.0307}} 
&\rm{318.633} &\rm{0.0408} &\rm{344.028} &\rm{0.0440} 
&\rm{279.038} &\rm{0.0366} &\rm{342.476} &\rm{0.0432} 
&\rm{357.522} &\rm{0.0474} 
\\
\hline

\multirow{4}{*}{\rotatebox{90}{\rm{Australia}}} & \rm{6}
&\pmb{\rm{41.395}} &\pmb{\rm{0.0401}} &\underline{\rm{48.015}} &\underline{\rm{0.0474}} 
&\rm{98.662} &\rm{0.0955} &\rm{73.516} &\rm{0.0706} 
&\rm{96.349} &\rm{0.0925} &\rm{63.071} &\rm{0.0641} 
&\rm{63.636} &\rm{0.0632} 

\\

\multirow{4}{*}{} & \rm{12} 
&\pmb{\rm{51.671}} &\pmb{\rm{0.0499}} &\underline{\rm{60.644}} &\underline{\rm{0.0593}} 
&\rm{89.865} &\rm{0.0876} &\rm{81.986} &\rm{0.0784} 
&\rm{102.031} &\rm{0.0978} &\rm{74.619} &\rm{0.0730} 
&\rm{84.339} &\rm{0.0840} 

\\

\multirow{4}{*}{} & \rm{24}
&\pmb{\rm{63.496}} &\pmb{\rm{0.0605}} &\underline{\rm{69.666}} &\underline{\rm{0.0666}} 
&\rm{79.967} &\rm{0.0762} &\rm{88.641} &\rm{0.0848} 
&\rm{111.410} &\rm{0.1045} &\rm{92.404} &\rm{0.0895} 
&\rm{116.626} &\rm{0.1216} 
\\
\multirow{4}{*}{} & \rm{48}
&\pmb{\rm{75.092}}
&\pmb{\rm{0.0719}} &\underline{\rm{79.998}} &\underline{\rm{0.0779}} 
&\rm{89.536} &\rm{0.0861} &\rm{93.839} &\rm{0.0905} 
&\rm{114.243} &\rm{0.1076} &\rm{101.363} &\rm{0.1013} 
&\rm{124.735} &\rm{0.1288} 
\\
\toprule
\end{tabular}
\label{exp1_table}
}
\end{table*}

\begin{figure*}[h!]
    \centering
    \includegraphics[width=\textwidth]{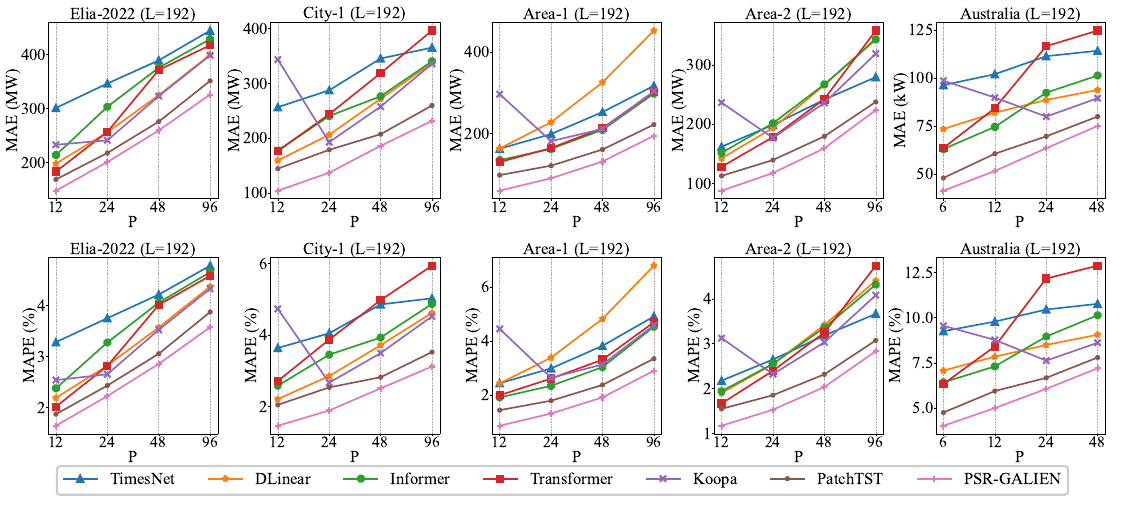}
    \caption{Forecasting results of each model with fixed input length (L=192) and varying forecasting length (P) on the Elia-2022, City-1, Area-1, Area-2 and Australia datasets, the prediction horizons in each dataset are 3h, 6h, 12h, and 24h in advance.}
    \label{exp1 picture}
\end{figure*}
In order to test the performance of each model in short-term power load forecasting scenarios, forecasting experiments with varying forecasting horizons of \{3h, 6h, 12h, 24h\} in advance are conducted on each dataset, and five experiments were conducted independently for each model with the parameter setup in Tab. \ref{hyper-table}, of which the best results are shown in Tab. \ref{exp1_table}, and the the prediction performance of each model with respect to the prediction length are shown in Fig. \ref{exp1 picture}.

As shown in the figure, the prediction error of each model has a different degree of upward trend as the forecasting length increases. Among them, PSR-GALIEN has the best forecasting accuracy under all conditions, with the PatchTST to be the second best, comparing with PatctTST, PSR-GALIEN shows substantial performance improvements at all prediction lengths, with average improvements of 22.7\%, 18.1\%, 10.4\%, and 8.6\%, this shows that the PSR-GALIEN model is particularly capable for intraday ultra-short-term load forecasting scenarios. Meanwhile, considering power loads with different characteristics, the model is robust as well. Taking the MAPE metrics as the indicator, in the day-ahead load forecasting scenarios which are important to the dispatch operation of the power system, the prediction accuracy of the PSR-GALIEN model for the city (City-1) and the regional distribution grid (Area-1, Area-2) can reach the level of about 97\%, and for the system-level load in the large-scale interconnected power system (Elia), the prediction accuracy can reach the level of about 96.5\%, and for the aggregated residential loads (Australia) with strong volatility can also reach a level of about 92.8\%. Overall, the PSR-GALIEN model achieves superior performance on these scenarios.

\subsubsection{Comparison experiments with varying input lengths}
\begin{figure}[h!]
    \centering
    \includegraphics[width=0.7\textwidth]{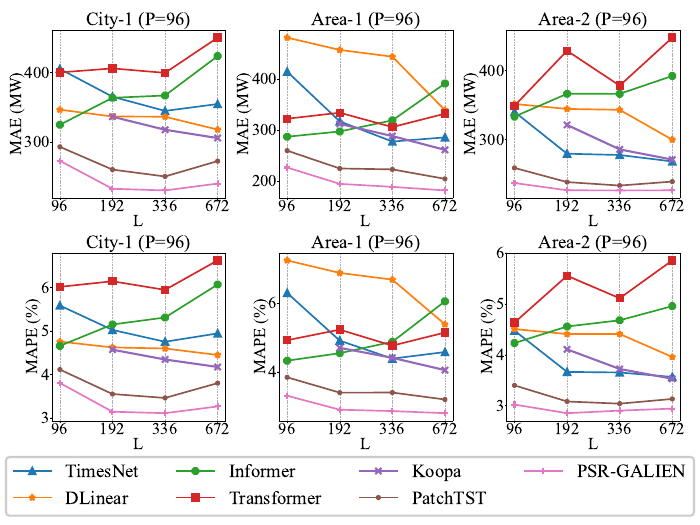}
    \caption{Forecasting results of each model with fixed out length (P=96) and varying input length (L) on the City-1, Area-1, Area-2 datasets.}
    \label{exp2_picture}
\end{figure}

The length of look-back window is an important parameter in models that adopt the autoregressive prediction strategy, and the increase of the window length brings more historical information, however, on the other hand, it also poses a challenge for the model to efficiently extract the useful features from the long input sequence. In general, a good deep learning model should be able to gain extra improvement in forecasting accuracy from longer inputs, however, the study in \cite{zeng2023transformers} argues that for the self-attention mechanism structure dominated models, they are inherently incapable of effectively utilizing long sequential data for prediction. Therefore, in order to validate these viewpoints and also to test the prediction performance of the PSR-GALIEN model for varying prediction lengths, in this part of the experiment, the output length of each model is fixed to P=96, and different input lengths of $\{96, 192, 336, 672\}$ steps are used for testing, the results are shown in Tab. \ref{exp2_table} and Fig. \ref{exp2_picture}.

\begin{table*}[h!]
\centering
\caption{Experimental results for different input lengths (L) on three long datasets under fixed forecasting length P=96, with the best \textbf{bolded}
and the second best \underline{underlined}. - represents the case that the relative experiments cannot be conducted with this parameter setup.}
\label{table1}
\resizebox{\linewidth}{!}{
\Large
\begin{tabular}{c|c|cc|cc|cc|cc|cc|cc|cc}
\toprule
\multicolumn{2}{c|}{\multirow{2}{*}{\rm{Model}}}
&\multicolumn{2}{c|}{\rm{PSR-GALIEN}}
&\multicolumn{2}{c|}{\rm{PatchTST} \cite{nie2022time}}
&\multicolumn{2}{c|}{\rm{Koopa} \cite{liu2024koopa}}
&\multicolumn{2}{c|}{\rm{DLinear} \cite{zeng2023transformers}}
&\multicolumn{2}{c|}{\rm{TimesNet} \cite{wu2022timesnet}}
&\multicolumn{2}{c|}{\rm{Informer} \cite{zhou2021informer}}
&\multicolumn{2}{c}{\rm{Transformer} \cite{vaswani2017attention}}

\\

\multicolumn{2}{c|}{} & \multicolumn{2}{c|}{\rm{Ours}} & \multicolumn{2}{c|}{\rm{2023}} & \multicolumn{2}{c|}{\rm{2024}} & \multicolumn{2}{c|}{\rm{2023}} & \multicolumn{2}{c|}{\rm{2023}} & \multicolumn{2}{c|}{\rm{2021}}& \multicolumn{2}{c}{\rm{2018}}\\
\hline
\multicolumn{2}{c|}{\rm{Metrics}} & \rm{MAE} & \rm{MAPE} &\rm{MAE} & \rm{MAPE} & \rm{MAE} & \rm{MAPE} & \rm{MAE} & \rm{MAPE} & \rm{MAE} & \rm{MAPE} & \rm{MAE} & \rm{MAPE} & \rm{MAE} & \rm{MAPE}\\ 
\hline

\multirow{4}{*}{\rotatebox{90}{\rm{City-1}}} & \rm{96}
&\pmb{\rm{273.254}} &\pmb{\rm{0.0380}} &\underline{\rm{293.259}} &\underline{\rm{0.0411}} 
&\rm{-} &\rm{-} &\rm{346.389} &\rm{0.0476} 
&\rm{404.786} &\rm{0.0559} &\rm{324.802} &\rm{0.0466} 
&\rm{399.749} &\rm{0.0602} 

\\

\multirow{4}{*}{} & \rm{192}
&\pmb{\rm{233.202}} &\pmb{\rm{0.0314}} &\underline{\rm{260.656}} &\underline{\rm{0.0355}} 
&\rm{336.351} &\rm{0.0457} &\rm{336.695} &\rm{0.0462} 
&\rm{365.023} &\rm{0.0502} &\rm{363.412} &\rm{0.0515} 
&\rm{405.517} &\rm{0.0615} 

\\

\multirow{4}{*}{} & \rm{336} 
&\pmb{\rm{230.821}} &\pmb{\rm{0.0311}} &\underline{\rm{250.805}} &\underline{\rm{0.0346}} 
&\rm{317.685} &\rm{0.0434} &\rm{336.335} &\rm{0.0460} 
&\rm{344.665} &\rm{0.0475} &\rm{366.725} &\rm{0.0531} 
&\rm{399.125} &\rm{0.0595} 

\\

\multirow{4}{*}{} & \rm{672}
&\pmb{\rm{240.650}} &\pmb{\rm{0.0326}} &\underline{\rm{272.837}} &\underline{\rm{0.0380}} 
&\rm{305.806} &\rm{0.0417} &\rm{317.891} &\rm{0.0445} 
&\rm{354.505} &\rm{0.0494} &\rm{423.107} &\rm{0.0607} 
&\rm{449.562} &\rm{0.0663} 

\\
\hline

\multirow{4}{*}{\rotatebox{90}{\rm{Area-1}}} & \rm{96} 
&\pmb{\rm{226.624}} &\pmb{\rm{0.0330}} &\underline{\rm{259.734}} &\underline{\rm{0.0384}} 
&\rm{-} &\rm{-} &\rm{481.462} &\rm{0.0724} 
&\rm{414.741} &\rm{0.0629} &\rm{287.090} &\rm{0.0432} 
&\rm{322.379} &\rm{0.0492} 

\\

\multirow{4}{*}{} & \rm{192} 
&\pmb{\rm{194.516}} &\pmb{\rm{0.0289}} &\underline{\rm{224.724}} &\underline{\rm{0.0339}} 
&\rm{312.171} &\rm{0.0470} &\rm{457.349} &\rm{0.0688} 
&\rm{317.123} &\rm{0.0491} &\rm{297.334} &\rm{0.0454} 
&\rm{334.147} &\rm{0.0523} 

\\

\multirow{4}{*}{} & \rm{336} 
&\pmb{\rm{188.676}} &\pmb{\rm{0.0286}} &\underline{\rm{222.945}} &\underline{\rm{0.0339}} 
&\rm{288.285} &\rm{0.0441} &\rm{444.073} &\rm{0.0668} 
&\rm{277.560} &\rm{0.0438} &\rm{319.182} &\rm{0.0487} 
&\rm{305.958} &\rm{0.0476} 

\\

\multirow{4}{*}{} & \rm{672} 
&\pmb{\rm{181.655}} &\pmb{\rm{0.0279}} &\underline{\rm{204.425}} &\underline{\rm{0.0319}} 
&\rm{261.266} &\rm{0.0405} &\rm{339.135} &\rm{0.0538} 
&\rm{285.750} &\rm{0.0457} &\rm{391.477} &\rm{0.0605} 
&\rm{332.299} &\rm{0.0514}  

\\
\hline

\multirow{4}{*}{\rotatebox{90}{\rm{Area-2}}} & \rm{96} 
&\pmb{\rm{236.410}} &\pmb{\rm{0.0302}} &\underline{\rm{258.376}} &\underline{\rm{0.0340}} 
&\rm{-} &\rm{-} &\rm{351.420} &\rm{0.0450} 
&\rm{339.944} &\rm{0.0447} &\rm{332.990} &\rm{0.0423} 
&\rm{348.711} &\rm{0.0463} 

\\

\multirow{4}{*}{} & \rm{192} 
&\pmb{\rm{225.869}} &\pmb{\rm{0.0285}} &\underline{\rm{237.778}} &\underline{\rm{0.0308}} 
&\rm{320.894} &\rm{0.0411} &\rm{344.183} &\rm{0.0441} 
&\rm{279.038} &\rm{0.0366} &\rm{366.199} &\rm{0.0455} 
&\rm{428.806} &\rm{0.0555} 

\\

\multirow{4}{*}{} & \rm{336} 
&\pmb{\rm{225.633}} &\pmb{\rm{0.0290}} &\underline{\rm{232.715}} &\underline{\rm{0.0304}} 
&\rm{285.305} &\rm{0.0372} &\rm{342.875} &\rm{0.0441} 
&\rm{277.149} &\rm{0.0365} &\rm{366.106} &\rm{0.0468} 
&\rm{378.030} &\rm{0.0511} 

\\

\multirow{4}{*}{} & \rm{672} 
&\pmb{\rm{225.812}} &\pmb{\rm{0.0294}} &\underline{\rm{238.517}} &\underline{\rm{0.0313}} 
&\rm{270.486} &\rm{0.0353} &\rm{299.436} &\rm{0.0396} 
&\rm{267.791} &\rm{0.0356} &\rm{392.164} &\rm{0.0495} 
&\rm{447.858} &\rm{0.0585} 

\\
\bottomrule
\end{tabular}
}
\label{exp2_table}
\end{table*}

From the experimental results, it is clear that the Transformer model along with its improved variant Informer do not benefit from a longer look-back window for their inefficient point-wise modeling approach, while the pure MLP-based Koopa and DLinear model, longer data means that a better fit can be made, therefore their forecasting performance is significantly improved. Finally, TimesNet, PatchTST and the PSR-GALIEN all show some improvement in prediction performance for longer input sequences, and the PSR-GALIEN remains the best, which should be attributed to the feature extraction methods used in it. TimesNet gains from longer inputs because of the additional frequency domain information from longer inputs, the 2D-CNN structure it employs is not able to model global features of the whole sequence well, which limits its prediction performance from further improvement. For PatchTST, its use of a fixed-length patch segmentation strategy may not necessarily be optimal for subsequent feature extraction as the input length changes. The PSR-GALIEN model can directly extract global and local features in the phase trajectories with varying length, thus combining the advantages of the above two models. 

\subsubsection{Ablation experiments}
\label{11}
\begin{table*}[h!]
\centering
\caption{The results of the ablation experiments on five datasets, where the input and output length settings used are the same as experiment one.}
\resizebox{\linewidth}{!}{
\Large
\begin{tabular}{c|c|rrrr|rrrrrrrr}
\toprule
\multicolumn{2}{c}{\rm{Dataset}} & \multicolumn{4}{c}{\rm{Elia-2022}} & \multicolumn{4}{c}{\rm{City-1}}    & \multicolumn{4}{c}{\rm{Area-1}} \\
\cline{3-14}    \multicolumn{2}{c}{\rm{Prediction Length}} & \multicolumn{1}{c}{\rm{96}} & \multicolumn{1}{c}{\rm{48}} & \multicolumn{1}{c}{\rm{24}} & \multicolumn{1}{c|}{\rm{12}} & \multicolumn{1}{c}{\rm{96}} & \multicolumn{1}{c}{\rm{48}} & \multicolumn{1}{c}{\rm{24}} & \multicolumn{1}{c|}{\rm{12}} & \multicolumn{1}{c}{\rm{96}} & \multicolumn{1}{c}{\rm{48}} & \multicolumn{1}{c}{\rm{24}} & \multicolumn{1}{c}{\rm{12}} \\
    \hline
    \rm{w/} & \rm{MAE}   & \rm{325.566}  & \rm{260.862}  & \rm{201.027}  & \rm{148.442}  & \rm{233.638}  & \rm{182.532}  & \rm{137.861}  & \multicolumn{1}{r|}{\rm{103.803} } & \rm{194.549}  & \rm{130.884}  & \rm{89.773}  & \rm{61.731}  \\
        \rm{Local Feature Extraction Module} & \rm{MAPE}  & \rm{0.0375}  & \rm{0.0289}  & \rm{0.0223}  & \rm{0.0164}  & \rm{0.0315}  & \rm{0.0245}  & \rm{0.0189}  & \multicolumn{1}{r|}{\rm{0.0145} } & \rm{0.0290}  & \rm{0.0190}  & \rm{0.0130}  & \rm{0.0089}  \\
    \hline
    \rm{w/o} & \rm{MAE}   & \rm{343.996}  & \rm{273.589}  & \rm{211.351}  & \rm{155.345}  & \rm{253.050}  & \rm{194.702}  & \rm{146.843}  & \multicolumn{1}{r|}{\rm{109.970}} & \rm{208.535}  & \rm{146.298}  & \rm{106.311}  & \rm{72.373}  \\
    \rm{Local Feature Extraction Module} & \rm{MAPE}  & \rm{0.0397}  & \rm{0.0302}  & \rm{0.0234}  & \rm{0.0172}  & \rm{0.0342}  & \rm{0.0263}  & \rm{0.0203}  & \multicolumn{1}{r|}{\rm{0.0154}} & \rm{0.0312}  & \rm{0.0215}  & \rm{0.0155}  & \rm{0.0106}  \\
    \hline
    \multicolumn{2}{c}{\rm{Dataset}} & \multicolumn{4}{c}{\rm{Area-2}}    & \multicolumn{4}{c}{\rm{Australia}} &       &       &       &  \\
\cline{3-10}    \multicolumn{2}{c}{\rm{Prediction Length}} & \multicolumn{1}{c}{\rm{96}} & \multicolumn{1}{c}{\rm{48}} & \multicolumn{1}{c}{\rm{24}} & \multicolumn{1}{c|}{\rm{12}} & \multicolumn{1}{c}{\rm{48}} & \multicolumn{1}{c}{\rm{24}} & \multicolumn{1}{c}{\rm{12}} & \multicolumn{1}{c}{\rm{6}} &       &       &       &  \\
\cline{1-10}    \rm{w/} & \rm{MAE}   & \rm{223.192}  & \rm{160.313}  & \rm{118.413}  & \rm{89.740}  & \rm{74.945}  & \rm{63.197}  & \rm{51.290}  & \rm{40.466}  &       &       &       &  \\
    \rm{Local Feature Extraction Module} & \rm{MAPE}  & \rm{0.0284}  & \rm{0.0205}  & \rm{0.0153}  & \rm{0.0119}  & \rm{0.0703}  & \rm{0.0600}  & \rm{0.0495}  & \rm{0.0393}  &       &       &       &  \\
\cline{1-10}    \rm{w/o} & \rm{MAE}   & \rm{233.803}  & \rm{171.153}  & \rm{127.105}  & \rm{94.569}  & \rm{80.881}  & \rm{69.807}  & \rm{54.723}  & \rm{42.686}  &       &       &       &  \\
        \rm{Local Feature Extraction Module} & \rm{MAPE}  & \rm{0.0297}  & \rm{0.0221}  & \rm{0.0167}  & \rm{0.0127}  & \rm{0.0759}  & \rm{0.0663}  & \rm{0.0529}  & \rm{0.0415}  &       &       &       &  \\
\bottomrule   
\end{tabular}%
}
\label{Abla-table}%
\end{table*}%


\begin{figure}[h!]
  \centering
\includegraphics[width=0.7\textwidth]{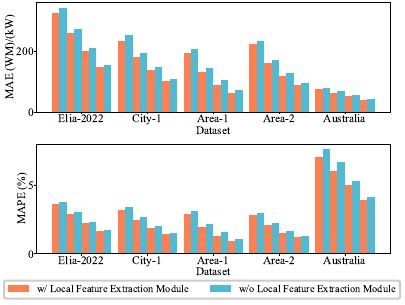}
\caption{Forecasting results of the PSR-GALIEN along with its variant without the local feaure extraction module}
\label{Abla_picture}
\end{figure}

In order to further explore the advantage of the proposed image-based forecasting approach for the extra utilization of local feature in the phase trajectory data, the ablation experiments are conduct to examine the impact of the local feature extraction module on model performance, it is worth noting that when the local feature extraction module is removed, this variant can be considered as a sequence-based forecasting model. It focuses on the prediction only using the information in the individual phase points which fails to consider the use of local temporal features in the projected sequence of the phase trajectory. For this purpose, this variant is used to conduct comparative experiments with the vanilla PSR-GALIEN model on five datasets, and the results are shown in Tab. \ref{Abla-table} and Fig. \ref{Abla_picture}.

From the results of the ablation experiments above, it can be seen that the local feature extraction module brings an average accuracy improvement of 5.9\%, 6.6\%, 12.1\%, 6.0\%, and 6.5\% on each dataset under the MAE metric respectively. On the one hand, this verifies the viewpoint in the previous sections that the preprocessing method of phase space reconstruction is equivalent to patch segmentation in terms of data structure, and the local temporal features under the perspective of the latter method can be extracted for the benefit of forecasting accuracy improvement. On the other hand, the 2D-CNN based local feature extraction module adopted in the PSR-GALIEN is well capable for the extraction of the corresponding features.

In the last subsection, we will continue to shed light on the issue of the learned global and local features during the forecasting process, and the feature visualisation techniques will be utilised to illustrate which area in the phase trajectory image are used by the model during the process of global and local feature extraction.

\subsubsection{Hyperparameters sensitivity experiments}

Given that the PSR-GALIEN model comprises multiple hyperparameters both in the preprocessing stage and within the network architecture, this section delves into a sensitivity analysis to evaluate their impact on forecasting performance.

\begin{figure*}[h!]
    \centering
    \includegraphics[width=\textwidth]{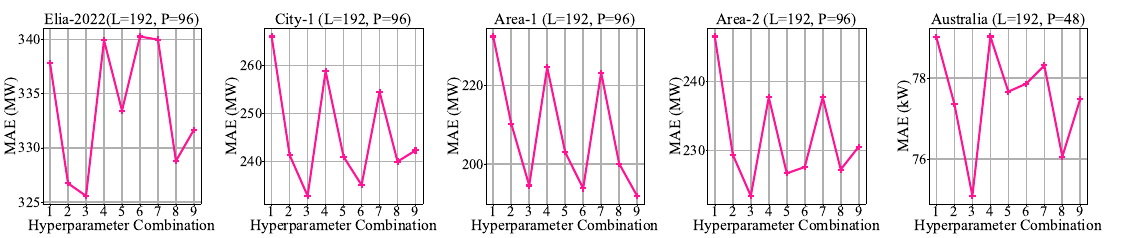}
    \caption{Forecasting performance of the PSR-GALIEN model with different model hyperparameter combinations in the five datasets}
    \label{hyper-picture}
\end{figure*}

First and foremost, the three most important hyperparameters in the network are $d_{\rm{model}}$, $d_{\rm{ff}}$, and $e_{\rm{layers}}$, as already shown in Fig.\ref{model-picture}. The candidate set of these three hyperparameters is set as follows:
$d_{\rm{model}}$=\{128, 256, 512\}, $d_{\rm{ff}}$ = 4$d_{\rm{model}}$, and $e_{\rm{layers}}$=\{2, 3, 4\}, which forms nine hyperparameter combinations. According to each of the above combinations, the PSR- GALIEN model is set up with the related hyperparameters for day-ahead forecasting on the five datasets with input length L = 192, and the results are shown in Tab. \ref{hyper-table} and Fig.\ref{hyper-picture}. From the experimental results, the hyperparameter combination 3 ($d_{\rm{model}}$=512, $d_{\rm{ff}}$=2048, $e_{\rm{layers}}$=2) has the best performance under the MAE evaluation metrics in all but Area-1 dataset, where its gap to the best one is only 1.5\%, thus the hyperparameter setup of combination 3 can be considered as the optimal choice for the PSR-GALIEN model.

\begin{figure*}[h!]
  \centering
\includegraphics[width=\textwidth]{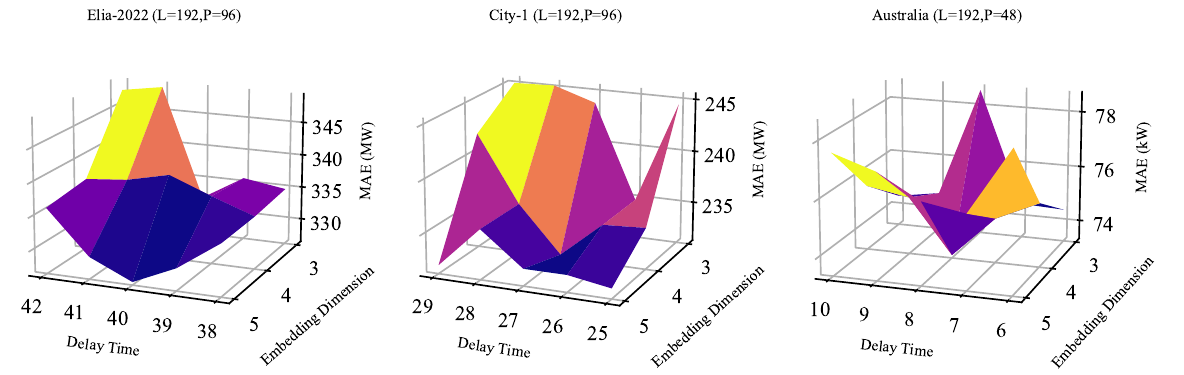}
\caption{Forecasting performance of the PSR-GALIEN model with different PSR parameter combinations in the three representative datasets: Elia-2022 (estimated optimal $\tau$=40, $m$=5), City-1 (estimated optimal $\tau$=27, $m$=4) and Australia (estimated optimal $\tau=8$, $m$=4) datasets.}
\label{Sensitivity-picture}
\end{figure*}

As the case with the preprocessing stage, considering that the choice of PSR parameters determines the extent to which the reconstructed phase trajectory approximates the inner attractor of the power load series, the Elia-2022, City-1 and Australia datasets are selected as the benchmark datasets in this part for their distinguished characteristics. The rest of the hyperparameters in the PSR-GALIEN model are set to the hyperparameter combination 3, with a input length of L=192 and an output length of P=96 (48), and the results of these experiments are shown in Tab. \ref{Sensitivity-tab} and Fig. \ref{Sensitivity-picture}. The results illustrate that the forecasting performance of the model remains generally stable for other parameters in the neighbourhood, with a low degree of accuracy degradation. Further exploring the two reconstruction parameters, it is clear that the embedding dimension $m$ has a greater impact on the prediction performance compared to the delay time $\tau$, this can be attributed to the fact that, as demonstrated in Eq. \ref{PSR embedding equa}, the embedding dimension $m$ has a greater effect on the length of the image obtained from PSR, especially in the datasets of Elia-2022, City-1, Area-1, Area-2, in which $\tau \gg m$. In such case, the dynamic features obtained in the global feature extraction process may not be able to retain the original characteristics of the attractor, thus degradation of forecasting performance occurs.
\begin{table*}[h!]
\centering
\caption{Experimental results on different model hyperparameter combinations in five datasets, with the input length L=192 and the PSR parameters fixed to that in Tab.\ref{dataset tab}, the best results are \textbf{bolded} and the second best are \underline{underlined}.}
\resizebox{0.6\linewidth}{!}{
\Large
\begin{tabular}{ccc|cccccr}
\toprule
\multicolumn{3}{c}{\rm{Dataset}} &       & \rm{Elia-2022} & \rm{City-1} & \rm{Area-1} & \rm{Area-2} & \ \rm{Australia} \\
\hline
\multicolumn{3}{c|}{\rm{Hyperparameters}}  & \rm{Preddiction}  & \multirow{2}[2]{*}{\rm{96}} & \multirow{2}[2]{*}{\rm{96}} & \multirow{2}[2]{*}{\rm{96}} & \multirow{2}[2]{*}{\rm{96}} & \multicolumn{1}{c}{\multirow{2}[2]{*}{\rm{48}}} \\
$d_{\rm{model}}$ & $d_{\rm{ff}}$   & $e_{\rm{layers}}$ & \rm{Length} &       &       &       &       &  \\
\hline
\multicolumn{3}{c|}{\rm{Combination 1}} & \rm{MAE}   & \rm{337.792}  & \rm{266.054}  & \rm{232.493}  & \rm{246.484}  & \rm{79.011}  \\
\rm{128}   & \rm{512}   & \rm{2}     & \rm{MAPE}  & \rm{0.0372}  & \multicolumn{1}{r}{\rm{0.0359}} & \multicolumn{1}{r}{\rm{0.0348} } & \multicolumn{1}{r}{\rm{0.0315}} & \rm{0.0743}  \\
\hline
\multicolumn{3}{c|}{\rm{Combination 2}} & \rm{MAE}   & \rm{\underline{326.725}}  & \rm{241.344}  & \rm{210.296}  & \rm{229.404}  & \rm{77.360}  \\
\rm{256}   & \rm{1024}  & \rm{2}     & \rm{MAPE}  & \rm{\underline{0.0360}}  & \rm{0.0327}  & \rm{0.0315}  & \rm{0.0292}  & \rm{0.0733}  \\
\hline
\multicolumn{3}{c|}{\rm{Combination 3}} & \rm{MAE}   & \multicolumn{1}{r}{\rm{\pmb{325.566}}} & \multicolumn{1}{r}{\rm{\pmb{232.814}}} & \multicolumn{1}{r}{\rm{194.516}} & \multicolumn{1}{r}{\rm{\pmb{223.444}}} & \rm{\pmb{75.092}}  \\
\rm{512}   & \rm{2048}  & \rm{2}     & \rm{MAPE}  & \rm{\pmb{0.0357}}  & \multicolumn{1}{r}{\rm{\pmb{0.0313}}} & \multicolumn{1}{r}{\rm{0.0289}} & \multicolumn{1}{r}{\rm{\pmb{0.0283}}} & \rm{\underline{0.0719}}  \\
\hline
\multicolumn{3}{c|}{\rm{Combination 4}} & \rm{MAE}   & \multicolumn{1}{r}{\rm{339.920}} & \multicolumn{1}{r}{\rm{258.819}} & \multicolumn{1}{r}{\rm{224.641}} & \multicolumn{1}{r}{\rm{237.704}} & \rm{79.030}  \\
\rm{128}   & \rm{512}   & \rm{3}     & \rm{MAPE}  & \rm{0.0373}  & \multicolumn{1}{r}{\rm{0.0347} } & \multicolumn{1}{r}{\rm{0.0336} } & \multicolumn{1}{r}{\rm{0.0303} } & \rm{0.0745}  \\
\hline
\multicolumn{3}{c|}{\rm{Combination 5}} & \rm{MAE}   & \multicolumn{1}{r}{\rm{333.434} } & \multicolumn{1}{r}{\rm{240.978} } & \multicolumn{1}{r}{\rm{203.008} } & \multicolumn{1}{r}{\rm{\underline{226.712}}} & \rm{77.668}  \\
\rm{256}   & \rm{1024}  & \rm{3}     & \rm{MAPE}  & \multicolumn{1}{r}{\rm{0.0367} } & \multicolumn{1}{r}{\rm{0.0323}} & \multicolumn{1}{r}{\rm{0.0303}} & \multicolumn{1}{r}{\rm{0.0290}} & \rm{0.0736}  \\
\hline
\multicolumn{3}{c|}{\rm{Combination 6}} & \rm{MAE}   & \multicolumn{1}{r}{\rm{340.259} } & \multicolumn{1}{r}{\rm{\underline{235.084}} } & \multicolumn{1}{r}{\rm{\underline{193.764}} } & \multicolumn{1}{r}{\rm{227.619} } & \rm{77.866}  \\
\rm{512}   & \rm{2048}  & \rm{3}     & \rm{MAPE}  & \multicolumn{1}{r}{\rm{0.0376}} & \multicolumn{1}{r}{\rm{\underline{0.0317}} } & \multicolumn{1}{r}{\rm{\underline{0.0287}} } & \multicolumn{1}{r}{\rm{0.0289} } & \rm{0.0744}  \\
\hline
\multicolumn{3}{c|}{\rm{Combination 7}} & \rm{MAE}   & \multicolumn{1}{r}{\rm{339.974}} & \multicolumn{1}{r}{\rm{254.525} } & \multicolumn{1}{r}{\rm{223.265} } & \multicolumn{1}{r}{\rm{237.677} } & \rm{78.316}  \\
\rm{128}   & \rm{512}   & \rm{4}     & \rm{MAPE}  & \multicolumn{1}{r}{\rm{0.0373} } & \multicolumn{1}{r}{\rm{0.0344} } & \multicolumn{1}{r}{\rm{0.0335} } & \multicolumn{1}{r}{\rm{0.0303} } & \rm{0.0736}  \\
\hline
\multicolumn{3}{c|}{\rm{Combination 8}} & \rm{MAE}   & \multicolumn{1}{r}{\rm{328.785}} & \multicolumn{1}{r}{\rm{239.979} } & \multicolumn{1}{r}{\rm{200.079} } & \multicolumn{1}{r}{\rm{227.253} } & \rm{\underline{76.063}}  \\
\rm{256}   & \rm{1024}  & \rm{4}     & \rm{MAPE}  & \multicolumn{1}{r}{\rm{0.0361} } & \multicolumn{1}{r}{\rm{0.0321} } & \multicolumn{1}{r}{\rm{0.0299 }} & \multicolumn{1}{r}{\rm{\underline{0.0288}} } & \rm{\pmb{0.0716}}  \\
\hline
\multicolumn{3}{c|}{\rm{Combination 9}} & \rm{MAE}   & \multicolumn{1}{r}{\rm{331.671}} & \multicolumn{1}{r}{\rm{242.292}} & \multicolumn{1}{r}{\rm{\pmb{191.850}}} & \multicolumn{1}{r}{\rm{230.555 }} & \rm{77.492}  \\
\rm{512}   & \rm{2048}  & \rm{4}     & \rm{MAPE}  & \rm{0.0365}  & \rm{0.0322}  & \rm{\pmb{0.0283}}  & \rm{0.0293}  & \rm{0.0729}  \\
\bottomrule
\end{tabular}%
}
\label{hyper-table}
\end{table*}

\begin{table}[h!]
    \centering
    \caption{Experimental results on different combinations of the PSR parameters in three dataset where significant difference in the delay time parameter exits. The other parameters in the experiment are set to L=192, P=96 (48), $d_{\rm{model}}=512$,
$d_{\rm{ff}}=2048$, $e_{\rm{layers}}=2$, the best results are \textbf{bolded}.}
\resizebox{0.6\linewidth}{!}{  
    \Large
    \begin{tabular}{c|c|rr|rr|rr}

\toprule
\multicolumn{2}{c|}{\rm{Embedding dimension}} & \multicolumn{2}{c|}{\rm{$m=3$}} & \multicolumn{2}{c|}{\rm{$m=4$}} & \multicolumn{2}{c}{\rm{$m=5$}} \\
\hline
\multicolumn{2}{c|}{\rm{Metrics}} & \multicolumn{1}{c}{\rm{MAE}} & \multicolumn{1}{c|}{\rm{MAPE}} & \multicolumn{1}{c}{\rm{MAE}} & \multicolumn{1}{c|}{\rm{MAPE}} & \multicolumn{1}{c}{\rm{MAE}} & \multicolumn{1}{c}{\rm{MAPE}} \\
\hline
\multirow{5}[2]{*}{\rotatebox{90}{\rm{Elia-2022}}}
& \rm{$\tau=38$}    & \rm{334.956}  & \rm{0.0369}  & \rm{334.350}  & \rm{0.0366}  &\rm{334.232}  & \rm{0.0367}  \\
& \rm{$\tau=39$}    & \rm{335.943}  & \rm{0.0367}  & \rm{336.633}  & \rm{0.0370}  & \rm{329.762}  & \rm{0.0363}  \\
& \rm{$\tau=40$}    & \rm{332.147}  & \rm{0.0365}  & \rm{339.109}  & \rm{0.0373}  & \pmb{\rm{326.875}}  & \pmb{\rm{0.0359}}  \\
& \rm{$\tau=41$}    & \rm{348.898} & \rm{0.0383}  & \rm{337.667}  & \rm{0.0373}  & \rm{329.749}  & \rm{0.0363}  \\
& \rm{$\tau=42$}    & \rm{347.881}  & \rm{0.0381 }& \rm{337.077}  & \rm{0.0372} & \rm{336.240}  & \rm{0.0368} \\
\hline
\multirow{5}[2]{*}{\rotatebox{90}{{\rm{City-1}}}} 
& \rm{$\tau=25$}    & \rm{244.677}  & \rm{0.0328}  & \rm{235.164}  & \rm{0.0315}  & \rm{232.199}  & \rm{0.0311}  \\
& \rm{$\tau=26$}   & \rm{245.408}  & \rm{0.0326}  & \rm{234.925}  & \rm{0.0313}  & \rm{232.845}  & \rm{0.0309}  \\
& \rm{$\tau=27$}    & \rm{243.995}  & \rm{0.0326}  & \pmb{\rm{231.051}}  & \pmb{\rm{0.0311}}  & \rm{232.837}  & \rm{0.0314}  \\
& \rm{$\tau=28$}    & \rm{245.282}  & \rm{0.0327}  & \rm{235.884}  & \rm{0.0319}  & \rm{235.966}  & \rm{0.0318}  \\
& \rm{$\tau=29$}    & \rm{245.154}  & \rm{0.0328} & \rm{242.166}  & \rm{0.0321} & \rm{232.027}  & \rm{0.0314} \\
\hline
\multirow{5}[2]{*}{\rotatebox{90}{{\rm{Australia}}}} 
& \rm{$\tau=6$}     & \rm{74.478}  & \rm{0.0706}  & \rm{75.599}  &\rm{0.0727}  & \rm{78.321}  & \rm{0.0733}  \\
& \rm{$\tau=7$}     & \rm{74.551}  & \rm{0.0703}  & \rm{74.901}  & \rm{0.0706}  & \rm{75.973}  & \rm{0.0714}  \\
& \rm{$\tau=8$}     & \rm{78.615}  & \rm{0.0744}  & \pmb{\rm{73.425}}  & \pmb{\rm{0.0691}}  & \rm{76.241}  & \rm{0.0723}  \\
& \rm{$\tau=9$}     & \rm{74.656}  & \rm{0.0710}  & \rm{75.359}  & \rm{0.0713}  & \rm{77.112}  & \rm{0.0735}  \\
& \rm{$\tau=10$}    & \rm{74.408}  & \rm{0.0706} & \rm{75.634}  & \rm{0.0710} & \rm{77.621}  & \rm{0.0730} \\
\bottomrule
\end{tabular}%
}
\label{Sensitivity-tab}
\end{table}

\subsection{Visualization analytics}
\subsubsection{Forecasting performance visualization}
\begin{figure}[h!]
\centering
\includegraphics[width=0.85\textwidth]{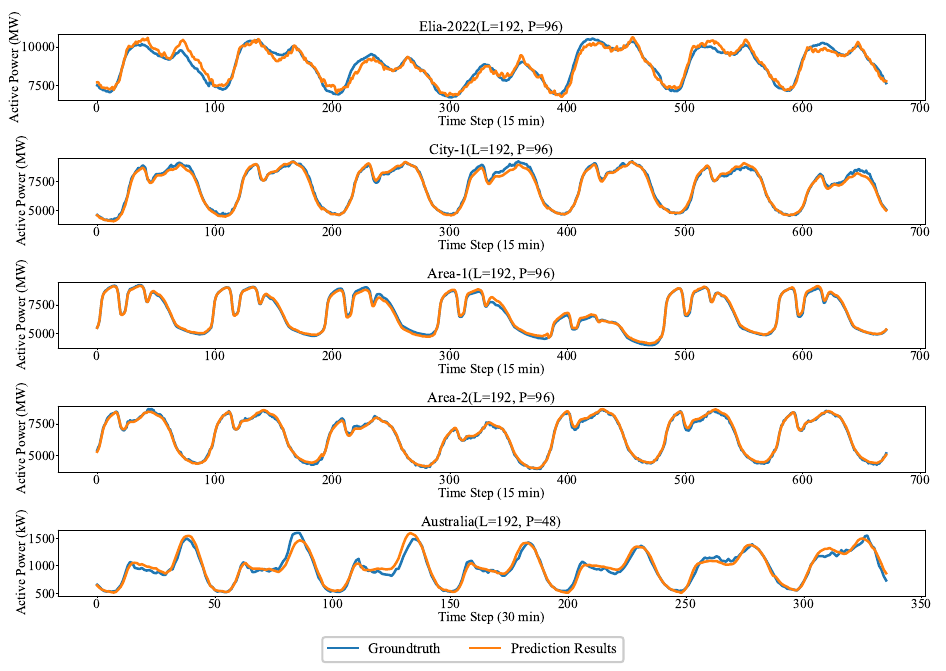}
\caption{Prediction results visualization}
\label{Prediction results visualization}
\end{figure}
The day-ahead forecasting performance of the PSR-GALIEN model on the five benchmark datasets are visualized in Fig. \ref{Prediction results visualization}, from the results, it is clear that the predictions results in the test datasets are quite close to the groundtruth values, which indicates that although the PSR-GALIEN model adopts the autoregressive forecasting scheme for the only usetilization of the historical power load series, its effective implementation of the PSR preprocessing method to uncover the chaotic characteristics of power loads in high-dimensional phase space, as well as the design of this global and local feature extraction strategy for feature extractionngineering, enable the model to be well capable in the day-ahead power load forecasting task among the power loads with varying temporal characteristics.

\subsubsection{Feature importance visualization}
\begin{figure*}[h!]
\centering
\includegraphics[width=\textwidth]{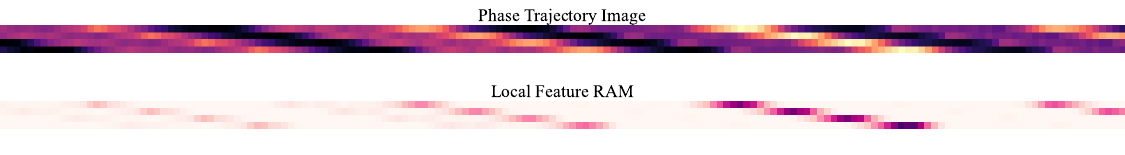}
\caption{Local feature visualization through RAM}
\label{Local feature}
\end{figure*}

\begin{figure*}[h!]
\centering
\includegraphics[width=\textwidth]{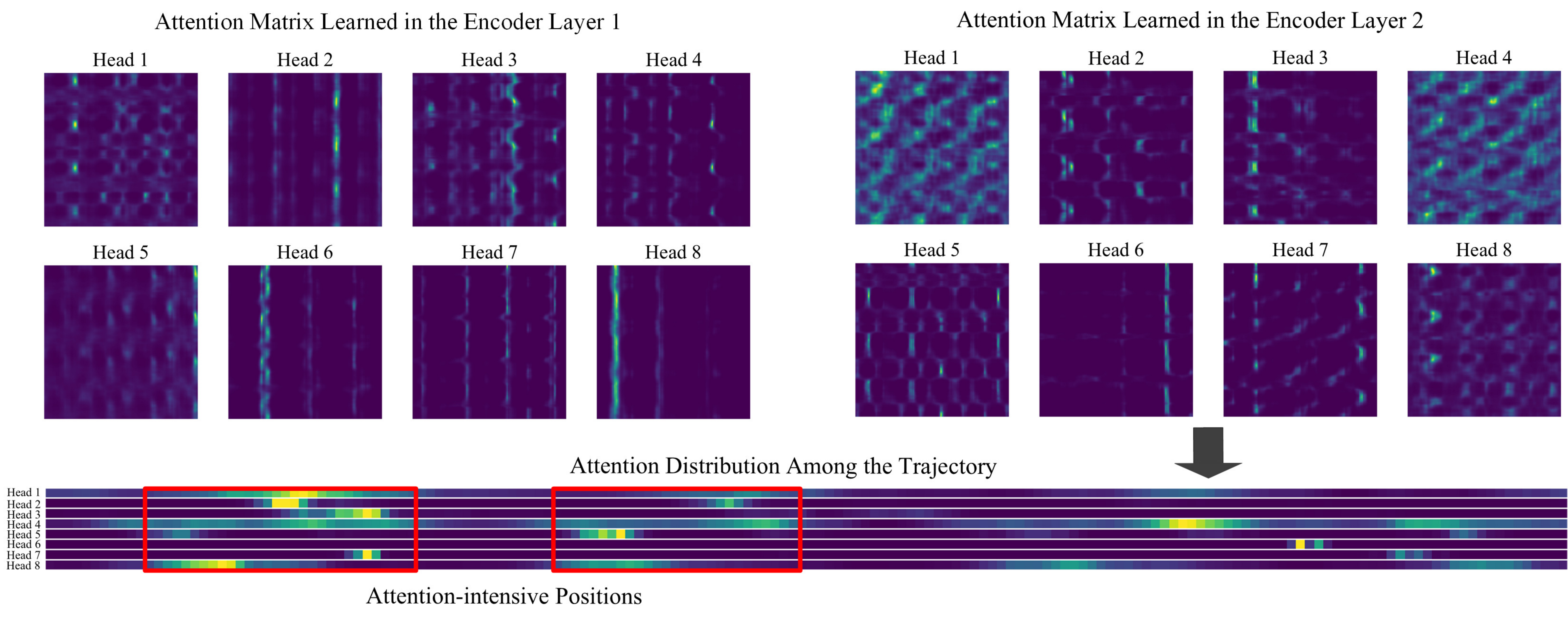}
\caption{Global feature visualization through heatmap}
\label{Global feature}
\end{figure*}
The focus of feature extraction in the PSR-GALIEN model is at the center of the discussion in this paper, in order to have an insight to these features, feature visualization techniques of heatmap and the regression activation map (RAM) \cite{XIE2023102771} are used in this subsection to show the corresponding features in the trajectory image utilized by the PSR-GALIEN model during the global and local feature extraction process.

RAM is a visualization technique that uses the gradient of back-propagation during inference process to reflect the extent to which specified regions in the feature maps of each CNN layer contribute to the final prediction results. In our study, the local patterns in the trajectory image noticed by the local feature extraction module can be obtained through the RAM in the last CNN layer of the local feature extraction module, which is shown in Fig.\ref{Local feature}. As we can see from the figure, the activated regions (the darker part of the image) in the RAM corresponds to the regions which have greater gradients, the darker the color of the activated regions in the RAM are, the more they contribute to the final prediction results. Comparing the RAM with the phase trajectory image, we can see that these activated regions are exactly the dark-colored regions of the original phase trace, which represent the the peak load periods in the original one-dimensional power load series. Thus, the conclusion can be made that the local patterns with high values in the projection subseries tend to be extracted by the local feature extraction module, and their contribution to the final forecasting result explains the phenomenon of the previous ablation experiments.

The heatmap is utilized to have a clear view of the distribution matrix learned in the self-attention process, in which shades of color can be used to represent the degree of dependence. The attention matrices of the two multi-head attention layers of the global feature extraction module are displayed in Fig. \ref{Global feature}, from which we can see, as the number of Encoder layer increases, more complex dependencies within input sequence (the phase trajectory sequence) begin to emerge. Thanks to the enhanced nonlinear characterization capabilities brought about by the MLP layer in each Transformer Encoder, deeper multi-head layer can model the nonlinear relationships which cannot be utilized in shallow layers. In the global feature extraction module, the Transformer Encoder is incorporated to obtain an effective nonlinear representation of the phase trajectory sequence for prediction, and the global features obtained come from the last point in the processed sequence. Therefore, from the attention matrix in the last Encoder layer, the contribution of each phase point (image patch) in the trajectory to the obtained global feature can be obtained and visualized, as demonstrated in the figure. In the multi-head attention mechanism, each head independently learns a dependency pattern in the sequence, from the results of the attention distribution in the sequences learned by each head, it is clear that some positions in the sequence are attention-intensive, which means greater attention are paid to these subseries when generating the global representation of the whole phase trajectory.

In summary, through the two feature visualization techniques, the process to learn both the global and local feature in the image can be clearly addressed, which in turn explains what part of the image's the model actually uses as a reference when making predictions.

\section{Conclusion}
\label{section 5}
In this study, firstly the relationship of the two preprocessing methods of PSR and PS in time series feature engineering is discussed, from the perspective of data structure, the equivalent relationship between them is demonstrated, which bridges a knowledge gap for the first time. Then, an image-based modeling approach for PSR with global and local feature extraction strategy is proposed, for the full utilization of this prior knowledge to extract useful patterns to improve the forecasting performance. On the basis of this method, an end-to-end deep learning model namely PSR-GALIEN for general multi-step forecasting is proposed, in which the Transformer Encoder and the 2D-CNN are implemented for the efficient extraction of these relative features. After that, the extensive experiments on five real datasets show that the proposed PSR-GALIEN model comprehensively outperforms the selected state-of-the-art models, thus its excellent performance as well as strong robustness in different power load forecasting scenarios are verified, at the same time, the effectiveness of this image-based engineering approach is validated, which provides a brand new perspective for the modeling and forecasting of power load series. Lastly, the RAM and heatmap are employed to visualized the utilized patterns in the trajectory image during the global and local feature extraction process, through which, the root cause of PSR-GALIEN's final prediction results  can be explained.

In the future, the multivariate modeling and forecasting approaches under the synergy of complex exogenous features will be further considered, in order to extend and generalize the proposed modeling approach as well as the PSR-GALIEN model in this paper.


\bibliographystyle{unsrt}
\bibliography{ref}

\end{document}